\documentclass[sigconf]{acmart}

\AtBeginDocument{%
  }
    
\usepackage{enumitem}
\usepackage{algorithm}
\usepackage{algpseudocode}
\usepackage{amsthm}
\usepackage{mathtools}
\usepackage{subcaption}
\usepackage{graphicx}
\usepackage{cancel}
\usepackage{tikz}
\usepackage{xcolor}
\usepackage{listings}
\usepackage{color}
\setcopyright{none}

\usepackage{amsmath}

\definecolor{checked}{RGB}{0, 0, 0}
\definecolor{dong}{RGB}{0, 0, 0}
\definecolor{camera}{RGB}{0,0,0}
\definecolor{fix}{RGB}{0,0,0}

\usepackage{xspace}

\copyrightyear{2026}
\acmYear{2026}
\setcopyright{cc}
\setcctype{by}
\acmConference[ICS '26]{2026 International Conference on Supercomputing}{July 06--09, 2026}{Belfast, United Kingdom}
\acmBooktitle{2026 International Conference on Supercomputing (ICS '26), July 06--09, 2026, Belfast, United Kingdom}
\acmDOI{10.1145/3797905.3807846}
\acmISBN{979-8-4007-2522-7/2026/07}

\begin{document}

\newcommand{\name}{CXL-CCL\xspace}
\newcommand{\nameone}{CXL-CCL-Aggregate\xspace}
\newcommand{\nametwo}{CXL-CCL-All\xspace}

\title[\name: Inter-Node Collective GPU-Communication Using a CXL Shared Memory Pool]{\name: Inter-Node Collective GPU-Communication Using a CXL Shared Memory Pool}
%Cross-server communication with shared CXL memory pool
%\subtitle{\normalsize{ICS 2026 Submission
    %\textbf{\#178} -- Confidential Draft -- Do NOT Distribute!!}}

% --- UC Merced ---
\author{Dong Xu}
\affiliation{
  \institution{University of California, Merced}
  \city{Merced}
  \state{California}
  \country{USA}
}
\email{dxu17@ucmerced.edu}

\author{Han Meng}
\affiliation{%
  \institution{University of California, Merced}
  \city{Merced}
  \state{California}
  \country{USA}
}
\email{hanmeng@ucmerced.edu}

% --- Zhejiang University ---
\author{Xinyu Chen}
\affiliation{%
  \institution{Zhejiang University}
  \city{Hangzhou}
   \country{China}
}
\email{xy.chen@zju.edu.cn}

% --- ByteDance ---
\author{Dengcheng Zhu}
\affiliation{%
  \institution{ByteDance}
  \city{San Jose}
  \state{California}
  \country{USA}
}
\email{dengcheng.zhu@bytedance.com}

\author{Wei Tang}
\affiliation{%
  \institution{ByteDance}
  \city{San Jose}
  \state{California}
  \country{USA}
}
\email{tangwei.0101@bytedance.com}

\author{Fei Liu}
\affiliation{%
  \institution{ByteDance}
  \city{San Jose}
  \state{California}
  \country{USA}
}
\email{fei.liu@bytedance.com}

\author{Liguang Xie}
\affiliation{%
  \institution{ByteDance}
  \city{Seattle}
  \state{Washington}
  \country{USA}
}
\email{liguang.xie@bytedance.com}

\author{Wu Xiang}
\affiliation{%
  \institution{ByteDance}
  \city{Beijing}
  \country{China}
}
\email{xiangwu@bytedance.com}

\author{Rui Shi}
\affiliation{%
  \institution{ByteDance}
  \city{Beijing}
  \country{China}
}
\email{shirui@bytedance.com}

\author{Yue Li}
\affiliation{%
  \institution{ByteDance}
  \city{San Jose}
  \state{California}
  \country{USA}
}
\email{theyueli@bytedance.com}

\author{Henry Hu}
\affiliation{%
  \institution{ByteDance}
  \city{San Jose}
  \state{California}
  \country{USA}
}
\email{henry.hu@bytedance.com}

\author{Hui Zhang}
\affiliation{%
  \institution{ByteDance}
  \city{San Jose}
  \state{California}
  \country{USA}
}
\email{hui.zhang@bytedance.com}

% --- Xconn-Tech ---
\author{Jianping Jiang}
\affiliation{%
  \institution{Xconn-Tech}
  \city{San Jose}
  \state{California}
    \country{USA}
}
\email{jp.jiang@xconn-tech.com}

% --- UC Merced (Last Author) ---
\author{Dong Li}
\affiliation{%
  \institution{University of California, Merced}
  \city{Merced}
  \state{California}
  \country{USA}
}
\email{dli35@ucmerced.edu}

\definecolor{dkgreen}{rgb}{0,0.6,0}
\definecolor{gray}{rgb}{0.5,0.5,0.5}
\definecolor{mauve}{rgb}{0.58,0,0.82}

\lstset{frame=single,
  language=Java,
  aboveskip=3mm,
  belowskip=3mm,
  showstringspaces=false,
  columns=flexible,
  basicstyle={\small\ttfamily},
  numbers=none,
  numberstyle=\tiny\color{gray},
  keywordstyle=\color{blue},
  commentstyle=\color{dkgreen},
  stringstyle=\color{mauve},
  breaklines=true,
  breakatwhitespace=true,
  tabsize=3,
 numbers=left,
  numberstyle=\scriptsize\color{gray},
  stepnumber=1,
  numbersep=4pt,
 linewidth=\columnwidth,      % force total width to fit the column
  numbers=left,
  numbersep=6pt,               % gap between numbers and code
  xleftmargin=2em,             % indent whole block
  framexleftmargin=2em 
}

%\author{\normalsize{ICS 2025 Submission
 %   \textbf{\#NaN} -- Confidential Draft -- Do NOT Distribute!!}}


\renewcommand{\shortauthors}{Dong Xu, et al.}


%Training large language models (LLMs) across multiple servers introduces significant pressure on GPU memory and interconnect bandwidth. Compute Express Link (CXL) memory pooling offers a scalable solution by enabling shared memory across hosts, reducing over-provisioning and improving resource utilization. We propose \name, a collective communication mechanism leveraging CXL shared memory to support cross-server GPU operations without relying on traditional RDMA-based networking. Our design includes a lightweight doorbell mechanism for synchronization, an interleaving data placement policy to utilize aggregate bandwidth, and a fine-grained chunk-based pipeline. Evaluation on a three-server system with a TITAN-II CXL switch and six Micron CZ120 memory cards, \name achieves efficient collective operations across hosts, demonstrating CXL's potential for scalable, memory-centric GPU communication. 

\begin{abstract}
Large language models (LLMs) training or inference across multiple nodes introduces significant pressure on GPU memory and interconnect bandwidth. The Compute Express Link (CXL) shared memory pool offers a scalable solution by enabling memory sharing across nodes, reducing over-provisioning and improving resource utilization. We propose \name, a collective communication library, leveraging the CXL shared memory pool to support cross-node GPU operations without relying on traditional RDMA-based networking. Our design addresses the challenges in synchronization, data interleaving, and communication parallelization faced by using the CXL shared memory pool for collective communications. Evaluating on multiple nodes with a TITAN-II CXL switch and six Micron CZ120 memory cards, we show that \name achieves highly efficient collective operations across hosts, demonstrating CXL's potential for scalable, memory-centric GPU communication. Our evaluation demonstrates that \name achieves average performance improvements of 1.34$\times$ for AllGather, 1.84$\times$ for Broadcast, 1.94$\times$ for Gather, and 1.07$\times$ for Scatter, compared to the original RDMA-based implementation over 200 Gbps InfiniBand. \textcolor{dong}{In addition, an LLM training case study shows 1.11$\times$ speedup compared with the InfiniBand while reducing interconnect hardware cost by 2.75$\times$.}  
\end{abstract}



\maketitle

\section{Introduction} 
%Large Language Models (LLMs) are rapidly increasing in size and complexity, leading to breakthroughs in natural language processing and generative AI.~\cite{} This growth places significant demands on hardware resources, including GPUs, high-bandwidth memory (HBM), CPU compute capacity, and system memory~\cite{}. This necessitates the use of multi-GPU setups distributed across multiple servers, where collective communication and synchronization between nodes become critical for maintaining throughput, reducing idle time, and ensuring overall system scalability.  
Large Language Models (LLMs) continue to grow in architectural complexity, driving major advances in natural language processing and generative AI \cite{bai2023qwen, team2023gemini, liu2024deepseek}. As a result, large‑scale training increasingly relies on multi‑GPU configurations distributed across multiple servers, where efficient collective communication and cross‑node synchronization are essential for sustaining throughput, minimizing idle time, and achieving scalable performance.

%In large-scale LLM training, collective communication is fundamental to maintaining efficiency and consistency across multiple GPUs and nodes. NVIDIA’s NCCL library provides optimized implementations of key collectives—such as all-reduce, broadcast, all-gather, and reduce-scatter—which are extensively used in both data-parallel and model-parallel training. For example, in data-parallel training, all-reduce synchronizes gradients after each backward pass, ensuring consistent model updates. In model-parallelism, all-gather and reduce-scatter help exchange activations or parameter shards between GPUs. Some architectures, such as Mixture of Experts (MoE), also introduce all-to-all communication to route and gather token batches across distributed expert layers. These communication patterns require high-throughput, low-latency collectives to avoid bottlenecks at scale. NCCL enables efficient use of high-speed interconnects like NVLink and InfiniBand, making it a critical component for scalable and performant LLM training across distributed systems. However, these technologies come with notable drawbacks, including high infrastructure cost, limited availability across cloud and on-premise environments, and complexity in software stack integration and tuning—especially when coordinating between NCCL, MPI, and various framework backends. These challenges must be carefully managed to fully realize the scalability potential of large distributed training systems.

In this distributed setting, collective communication plays a central role in maintaining both efficiency and consistency across GPUs and nodes \cite{deepspeed-megatron-lm, si2026collectivecommunication100kgpus, cclddp,fsdp, NCCL}. NVIDIA’s NCCL library~\cite{NCCL} provides highly optimized implementations of key collectives—such as all‑reduce, broadcast, all‑gather, and reduce‑scatter—which are heavily used in data‑parallel and model‑parallel training \textcolor{dong}{and inference}. For instance, data‑parallel training depends on all‑reduce to synchronize gradients after each backward pass, while model parallel approaches rely on all‑gather and reduce‑scatter to exchange activations or parameter shards. Architectures like Mixture of Experts (MoE) \cite{mu2026comprehensivesurveymixtureofexpertsalgorithms,MoE,Cai_2025} further introduce all‑to‑all communication to route and aggregate token batches across distributed expert layers. These communication patterns demand high‑bandwidth, low‑latency communication to avoid bottlenecks at scale. NCCL enables efficient use of high‑speed interconnects such as NVLink and InfiniBand~\cite{nvidia2021infiniband}, making it a foundational component of scalable LLM training \textcolor{dong}{and inference}. However, these technologies also introduce challenges, including high infrastructure cost, limited availability across cloud and on‑premise environments, and increased software‑stack complexity—particularly when coordinating NCCL, MPI, and framework‑level backend. Effectively managing these challenges is crucial to unlocking the full scalability potential of large distributed training \textcolor{dong}{and inference} systems.

% Collective communications are widely used in current LLM training framworks. Such like 
% To build a efficient communication network, existing communication hardwares such like Infiniband, NVLink provides huge bandwidths for servers. However, to enable high bandwidth between servers, it could easily cost millions to build a high bandwidth network between servers.
% \textcolor{fix}{Need to add more content to make it fluent}

%The emergence of the Compute Express Link (CXL) standard presents an opportunity to design high-performance memory pools. By providing a direct, low-latency, load/store memory interface, CXL allows CPU/GPUs to access remote memory with an efficiency that approaches local memory access. For example, Beluga and TraCT use the shared CXL memory pool to store the generated KV cache. The decoding servers can fetch the needed data from this shared KV pool.  These works are using this memory pool as a shared long-term storage. 

The emergence of the Compute Express Link (CXL)~\cite{cxl_website} creates opportunities for building high‑performance, shared memory pools. By offering a direct, low‑latency load/store interface, CXL enables CPUs and GPUs to access remote memory with efficiency approaching that of local memory. Recent systems such as Beluga~\cite{yang2025belugacxlbasedmemoryarchitecture} and TraCT~\cite{yoon2025tractdisaggregatedllmserving} leverage the CXL shared memory pool to store generated KV caches in LLM, allowing decoding servers to fetch required data directly from the pool. In these designs, the CXL memory pool primarily serves as a shared long‑term storage layer.

%In this paper, we contend that employing a shared CXL memory pool for collective communication represents an additional, and particularly important, use case. The shared CXL memory pool not only supplies a unified memory space across multiple servers, but also provides an interconnect fabric among servers and GPUs. By leveraging the shared CXL memory pool, applications are relieved from the need to maintain a complex, layered software stack. Owing to the memory semantics defined by the CXL protocol, different collective operations can directly write data into the memory pool and subsequently read the required data from it, thereby simplifying data movement and coordination across participating nodes. Besides the data access, current CXL memory pool can support up to 64GB/s bandwidth for each Servers depends on how many links it can connect to the CXL switch.

In this paper, we argue that using a CXL shared memory pool for collective communication represents an additional—and high-performance—use case. Beyond providing a unified memory space across nodes, the CXL shared memory pool effectively acts as an interconnect fabric among nodes and GPUs. By exploiting memory semantics defined in the CXL protocol, collective operations can write data directly into the pool and later read the required data without relying on a complex, layered communication stack. This simplifies data movement and coordination across nodes. 

% Moreover, current CXL memory pools can deliver up to 64 GB/s of bandwidth per node, depending on the number of links connected to the CXL switch. \textcolor{red}{(what conclusion do you want to draw from the above sentence?)}

%However, several challenges arise when designing efficient collective communication mechanisms based on a shared CXL memory pool. First, cross-node atomic lock operations are not supported. As a result, it is not possible to establish a critical section across nodes. When a collective operation is initiated, each rank is unable to determine when the required data becomes available in the shared memory, which can compromise correctness. Furthermore, incorrect assumptions about data readiness can lead to suboptimal timing of read operations and, consequently, performance inefficiencies. Second, current hosts do not support fine-grained interleaving at the cache-line level, as is typical for DRAM. Without careful data placement for each rank, concurrent memory accesses that target the same device will fail to fully exploit the aggregate bandwidth provided by the CXL switch.  Third, we observe that, for each rank, data transmission and reception are mutually independent operations. The only dependency is that a receive operation requires the corresponding data to be available in the shared memory pool. A fine-grained partitioning of the data is required to enable concurrent and overlapping data write and read operations.

However, designing efficient collective communication mechanisms atop a CXL shared memory pool introduces three challenges.  \textit{First}, current nodes do not support fine‑grained cache‑line interleaving in the CXL shared memory pool as in DRAM. %Without careful data placement,
As a result, concurrent accesses from multiple ranks may contend for the same CXL device, preventing full utilization of the aggregate bandwidth offered by the CXL switch~\cite{xconn}. \textcolor{dong}{\textit{Second}, 
the paradigm of shared memory pool for communication naturally introduces dependency between the sender and receiver ranks, which limits communication and computation parallelisms we can leverage to improve performance.} \textit{Third}, \textcolor{dong}{there is limited support for cross-node synchronization in the CXL shared memory pool, making it difficult to establish communication coordination. As a result, when a collective operation begins, each communication rank lacks a reliable method to determine data availability, causing incorrect behavior and poorly timed reads.} %when the necessary data becomes available in shared memory, risking incorrect behavior and poorly timed reads.

%data transmission and reception for each rank are independent operations, with the only dependency being that a receive must wait for its corresponding data to appear in the pool. Achieving high concurrency therefore requires fine‑grained data partitioning to enable overlapping writes and reads.

%To address these challenges, we propose \name, a collective communication library built on a CXL-based shared memory pool. First, \name introduces a lightweight in-memory pool locking mechanism, referred to as the doorbell mechanism. \name employs a computation-based doorbell allocation strategy to eliminate the need for maintaining heavyweight metadata and to reduce multiple memory pool accesses for lock acquisition. Second, \name presents two software interleaving techniques designed to fully exploit the bandwidth provided by the CXL switch. Third, \name enables fine-grained data overlapping between dependent write and read operations on the memory pool.

To address the above challenges, we propose \name, a collective communication library built for the CXL shared memory pool. First, \name incorporates two software‑level interleaving techniques, aiming to maximize utilization of the bandwidth provided by the CXL switch. Second, \textcolor{dong}{\name enables fine‑grained data partitioning, and hence enables overlap between dependent write and read operations within the memory pool}, allowing communication tasks to proceed with higher concurrency and reduced idle time. Third, \name introduces a lightweight in‑memory locking mechanism, referred to as the doorbell mechanism. By using a computation‑driven doorbell allocation strategy, \name avoids maintaining heavyweight metadata and reduces redundant memory‑pool accesses during lock acquisition.

The major contributions of this paper are summarized as follows:
\begin{itemize}
    \item To the best of our knowledge, this is the first work to use a CXL shared memory pool to perform collective communication across GPUs located on different nodes. \textcolor{dong}{This work introduces a fundamentally new method for high-performance collective communication.} 
    
    \item We develop \name, a collective communication library for GPUs that leverages a realistic CXL‑based shared memory pool. \textcolor{dong}{We characterize the performance of this new memory architecture from the perspective of memory pool-based communication, which is unprecedented.}
    
    %Within \name, we introduce three key mechanisms:(i) two software‑level interleaving techniques that maximize utilization of the internal bandwidth across diverse collective primitives;  (ii) fine‑grained overlap of write and read operations to improve overall communication efficiency; and (iii) a lightweight locking scheme that enforces correct access ordering among ranks during collective operations; 
    
    \item We evaluate \name on a multi-node GPU cluster. %where each node is equipped with an NVIDIA H100 GPU. 
    Our evaluation results indicate that \name achieves  1.34$\times$ speedup for AllGather, 1.84$\times$ for Broadcast, 1.94$\times$ for Gather, 1.07$\times$ for Scatter, 1.5$\times$ for AllReduce, 1.43$\times$ for ReduceScatter, 1.70$\times$ for Reduce, and 1.53$\times$ for AlltoAll, averaged over varying message sizes, compared to the original RDMA-based implementation over 200 Gbps InfiniBand. Our LLM training case study shows 
    %fof training a Llama-3-8B model with Wikipedia dataset using FSDP on multiple nodes, shows that we can get 
    1.11$\times$ speedup compared with using RDMA-based implementation over 200 Gbps InfiniBand, \textcolor{dong}{while reducing interconnect hardware cost by 2.75$\times$..}  
\end{itemize}

\section{Background} 
\label{sec:background}
\subsection{GPU Collective Communication Library: NCCL}
%NVIDIA Collective Communications Library (NCCL) is a library for highly optimized GPU-to-GPU communication. It is essential for scaling AI and HPC applications across multiple GPUs and nodes by accelerating collective operations like AllReduce, Broadcast, and Reduce, using topology-aware algorithms for high bandwidth and low latency over interconnects like NVLink, PCIe, and Ethernet. It provides low-level primitives for deep learning frameworks (like PyTorch and TensorFlow) to manage data movement efficiently, making large-scale distributed training feasible and performant.

NVIDIA Collective Communications Library (NCCL) is a high performance library designed for optimized GPU‑to‑GPU communication. It plays a critical role in scaling AI and HPC workloads across multiple GPUs and nodes by accelerating collective \textcolor{dong}{communication primitives} such as AllReduce, Broadcast, and Reduce. NCCL uses topology‑aware algorithms to deliver high bandwidth and low latency over interconnects including NVLink, PCIe, and Ethernet. By exposing efficient low‑level communication primitives, NCCL enables deep learning frameworks like PyTorch~\cite{pytorch_website} and TensorFlow~\cite{tensorflow_website} to orchestrate data movement effectively, making large‑scale distributed training both feasible and highly performant. %Table~\ref{tab:nccl_primitives} presents the set of collective communication primitives together with a concise description of each operation. 
We use the terms ``node'' and ``server'' interchangeably in the rest of the paper.

\subsection{Compute Express Link (CXL)}

%CXL is an emerging interconnect protocol designed to facilitate efficient memory sharing between host CPUs and accelerators. It defines three distinct protocols: CXL.IO, dedicated to CXL device management and DMA data transfers; CXL.cache, for coherent device access to host memory; and CXL.mem, enabling host load/store access to device-attached memory for expansion and pooling. The latest standard has enhanced system capabilities to enable efficient memory sharing and pooling by switching and cache coherency support.

CXL is an emerging interconnect protocol designed to enable efficient memory sharing between host CPUs and accelerators. It encompasses three complementary sub‑protocols: \texttt{CXL.io}, which handles device management and DMA transfers; \texttt{CXL.cache}, which provides coherent device accesses to host memory; and \texttt{CXL.mem}, which allows the host to perform load/store operations on device‑attached memory for expansion and pooling. The CXL specification 3.0 further strengthens these capabilities by introducing switching and enhanced cache coherency, making large‑scale memory sharing and pooling more efficient and practical. 

%This paper focuses primarily on memory expansion solutions leveraging the CXL.mem interface. Currently, the design and production of hardware supporting CXL.mem remains in its early stages. Most existing hardware only supports CXL 1.1, providing memory expansion capabilities for single servers. Although several research efforts have developed FPGA-based prototypes supporting CXL 2.0, their capabilities are constrained in both device count and memory capacity [41–45]. XConn [25, 46] has produced the first commercial CXL 2.0 switch, providing CXL.mem interfaces with support for 256 lanes and concurrent access from multiple hosts, delivering minimal 64-byte I/O latency of \textcolor{fix}{~750ns} and maximum switching capacity of 2TB/s. The introduction of CXL 2.0 switches fundamentally elevates the protocol beyond the single-node memory expansion of CXL 1.1, thereby enabling the creation of a large-scale shared memory pool.

This paper focuses on memory expansion solutions built on the \texttt{CXL.mem} sub-protocol. At present, the hardware design and production for \texttt{CXL.mem} remain in an early stage. Most commercially available devices support only CXL 2.0 which enables memory expansion within a single server. Xconn-tech~\cite{xconn}  has introduced the first commercial CXL 2.0 switch, offering \texttt{CXL.mem} interfaces with 256 lanes, multi‑host concurrent access, a minimal 64‑byte I/O latency of \textcolor{fix}{~658ns}, and a maximum switching bandwidth of 2 TB/s. The emergence of CXL 2.0 switches marks a significant step forward, extending the protocol beyond the single‑node memory expansion model in CXL 1.1 and enabling the construction of large‑scale shared memory pools.

\begin{figure}
    \centering
    \includegraphics[width=1.05\linewidth]{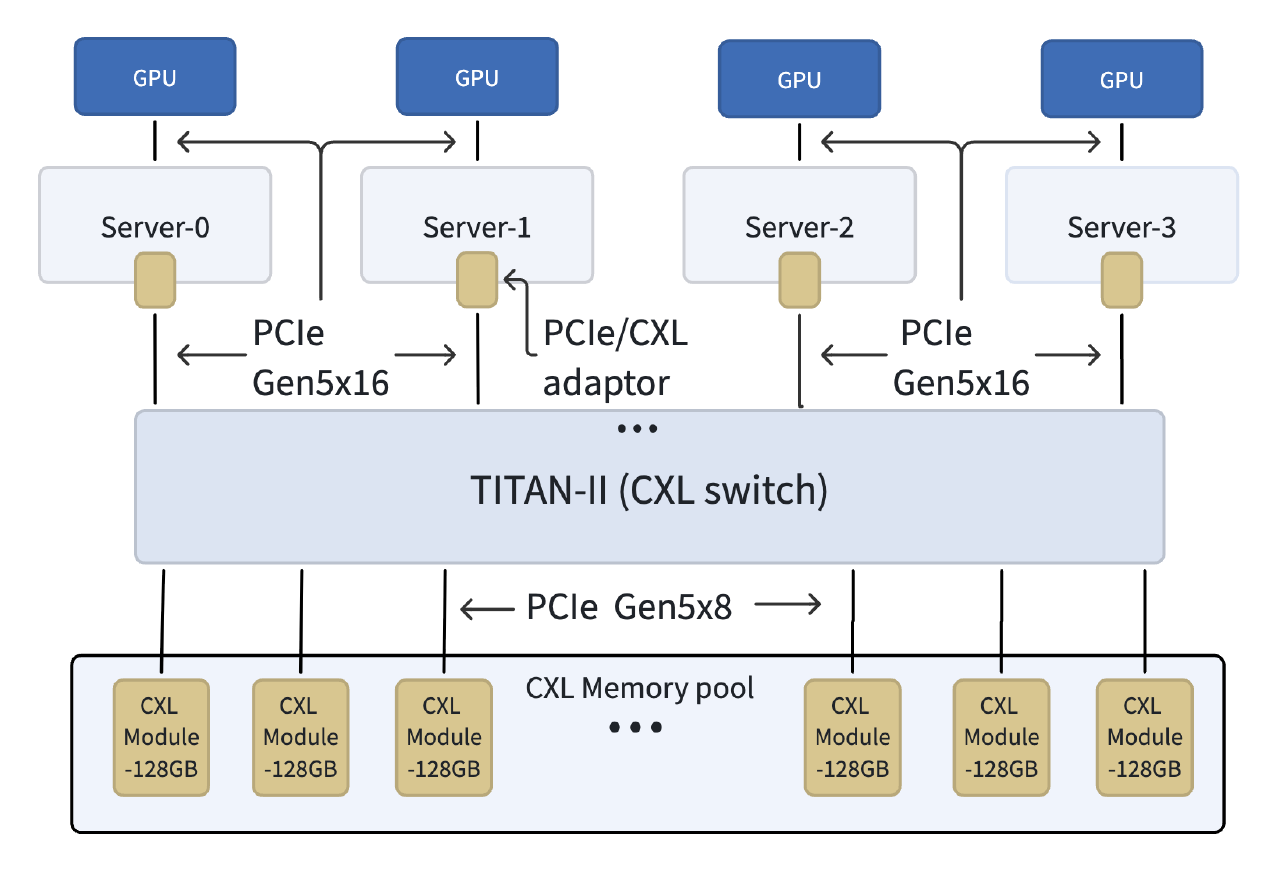}
    \caption{Architecture of the CXL shared memory pool.}
    \label{fig:architecure}
\end{figure}

Figure~\ref{fig:architecure} illustrates our system for study. All servers in the figure are connected to an Xconn-tech CXL switch through a Gen5x16 cable. Besides the servers, there is a memory-pool box encapsulating multiple CXL cards. Each card is connected to the switch via a Gen5x8/Gen5x16 cable.
 
%\textcolor{fix}{\textbf{Sharing through DAX.}} A CXL shared memory pool can be used as a memory resource that is accessible from multiple servers. However, current Linux does not provide a native NUMA mode that allows multiple hosts to share the same CXL memory as normal system memory. As a result, the shared pool is often exposed through Direct Access (DAX) as a device that applications can map into their address space. With DAX, processes access the shared CXL region with load/store semantics and bypass the page cache, which makes it a practical approach for sharing CXL memory across machines under current OS support limits. 

\textcolor{checked}{\textbf{Sharing through DAX.}} A CXL shared memory pool can serve as a memory resource accessible to multiple servers. However, current Linux kernels do not provide a native NUMA mode that allows multiple hosts to treat shared CXL memory as standard system memory. Consequently, the shared pool is typically exposed through Direct Access (DAX), allowing applications to map it directly into their address space. Using DAX, the processes access the shared CXL region with load/store semantics while bypassing the page cache, making it a practical mechanism for inter-node CXL memory sharing given the limitation of current operating system. %%%(OS) support.

%\textcolor{fix}{\textbf{Sequential stacked memory address.}} A common way to scale a CXL memory pool is to stack multiple CXL memory cards so their capacity adds up. In this setup, each card contributes a fixed amount of memory, and the system exposes the total pool as one larger address space. Sequential stacking means the total address space is built by placing each device’s memory right after the previous one, in a fixed order. For example, if there are 6 CXL devices and each provides 128 GB, the pool size is 768 GB: the address range [0, 128 GB) is backed by device-0, [128 GB, 256 GB) by device-1, [256 GB, 384 GB) by device-2, and this pattern continues until device-5 covers [640 GB, 768 GB).

\begin{figure}[!t]
    \centering
    \includegraphics[width=1\linewidth]{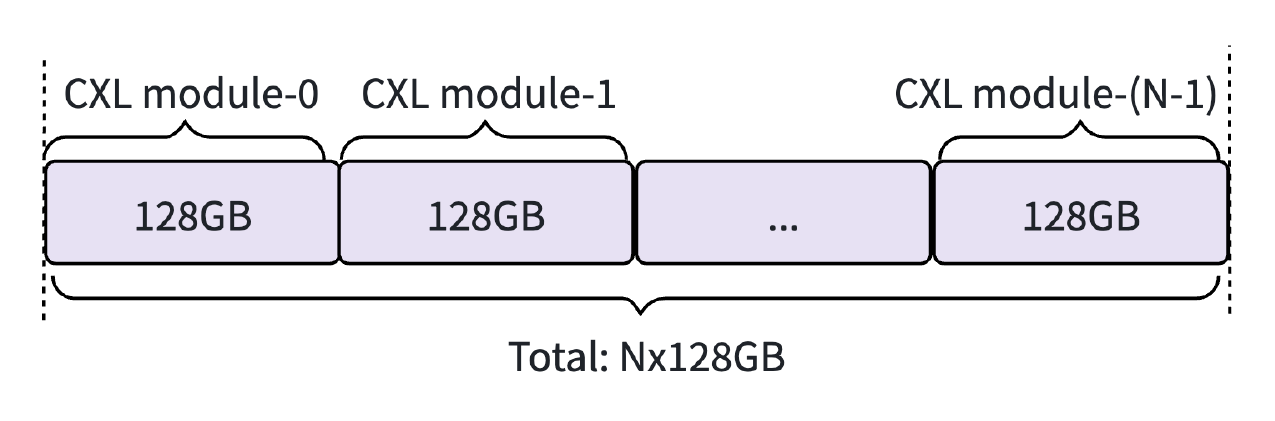}
    \caption{Sequentially stacked memory address space.}
    \label{fig:memory_address}
    \vspace{-10pt}
\end{figure}

\textcolor{checked}{\textbf{Sequentially stacked memory addresses.}} A common approach to scaling a CXL memory pool is to stack multiple CXL memory cards so that their capacities accumulate. In this configuration, each card contributes a fixed amount of memory capacity, and the system exposes the combined capacity as a single, contiguous address space. Sequential stacking builds the pool by placing each device’s memory range directly after another one in a predetermined order. For instance, with six CXL devices each providing 128 GB, the total pool size becomes 768 GB: the addresses [0, 128 GB) map to device 0, [128 GB, 256 GB) to device 1, and so on, until [640 GB, 768 GB) to device 5, shown in Figure~\ref{fig:memory_address}.

\begin{lstlisting}[language=C++, caption={Pseudo code for mapping a CXL pool via DAX and memory pinning for GPU DMA}, label={lst:cxl_dax_pin}, float=t]
fd = open("/dev/dax0.0", O_RDWR);
base = mmap(NULL, size, PROT_READ | PROT_WRITE, MAP_SHARED,fd, offset);//mapping           

ptr  = (uint8_t*) base;    // raw byte-address space
addr = ptr + alloc_offset;// manual allocation 
addr[0] = value;          // CPU store
x = addr[0];              // CPU load

cudaHostRegister(base, size, cudaHostRegisterDefault);  // pin (page-lock) for GPU DMA

cudaMemcpyAsync(gpu_buf, addr, bytes, cudaMemcpyDefault, stream);//CXL<->GPU example

munmap(base, size);
close(fd);
\end{lstlisting}

\textcolor{checked}{\textbf{Accessing the CXL shared memory pool}} %Accessing the CXL shared memory pool 
follows a workflow, illustrated in Listing  \ref{lst:cxl_dax_pin}. The listing shows how a host process uses the DAX (Device‑DAX) interface to map the shared pool and prepare the region for GPU DMA transfers. At Line 1, the process opens the DAX character device (e.g., /dev/dax0.0) with read/write permissions, treating the shared pool as a file‐like resource. Line 2 maps a \textit{size}-byte window of the device to the virtual address space of the process through mmap, producing a CPU‑accessible pointer (``base'') at an aligned offset. After mapping, the code performs manual layout management within the region: it casts ``base'' to a raw byte pointer (Line 4) and computes an application‑specific address (``addr'') by adding an allocation offset (Line 5). The host can then issue memory loads and stores to the shared pool (Lines 6–7), effectively reading and writing shared data buffers. To support efficient GPU accesses, the mapped pages are pinned using ``cudaHostRegister'' (Line 9), ensuring they are page‑locked and suitable for high‑throughput DMA. The host can then launch asynchronous transfers between the GPU buffer and the shared‑pool address using ``cudaMemcpyAsync'' (Line 11), enabling GPU‑driven communication through the shared memory pool. Once the memory operations complete, the process unmaps the region with ``munmap'' (Line 13) and closes the device handle (Line 14).

\section{Performance Characteristics of the CXL Shared Memory Pool}\label{sec:characteritics}

\begin{figure*}[!t]
\centering

\begin{subfigure}[t]{0.32\textwidth}
    \centering
    \includegraphics[width=\linewidth]{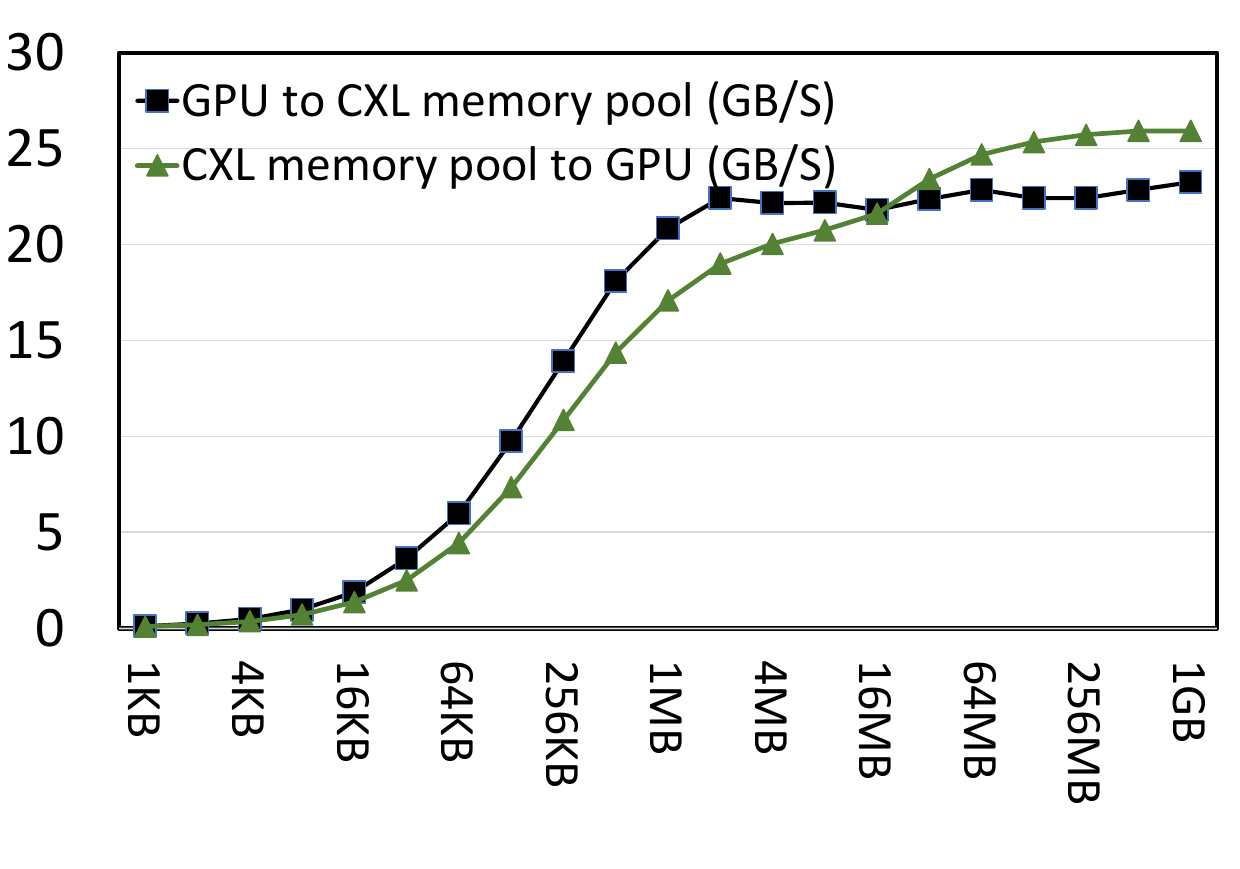}
    \caption{Data transfer bandwidth, evaluated with single node-exclusive accesses.}
    \label{fig:gpucxlbw}
\end{subfigure}
\hfill
\begin{subfigure}[t]{0.32\textwidth}
    \centering
    \includegraphics[width=\linewidth]{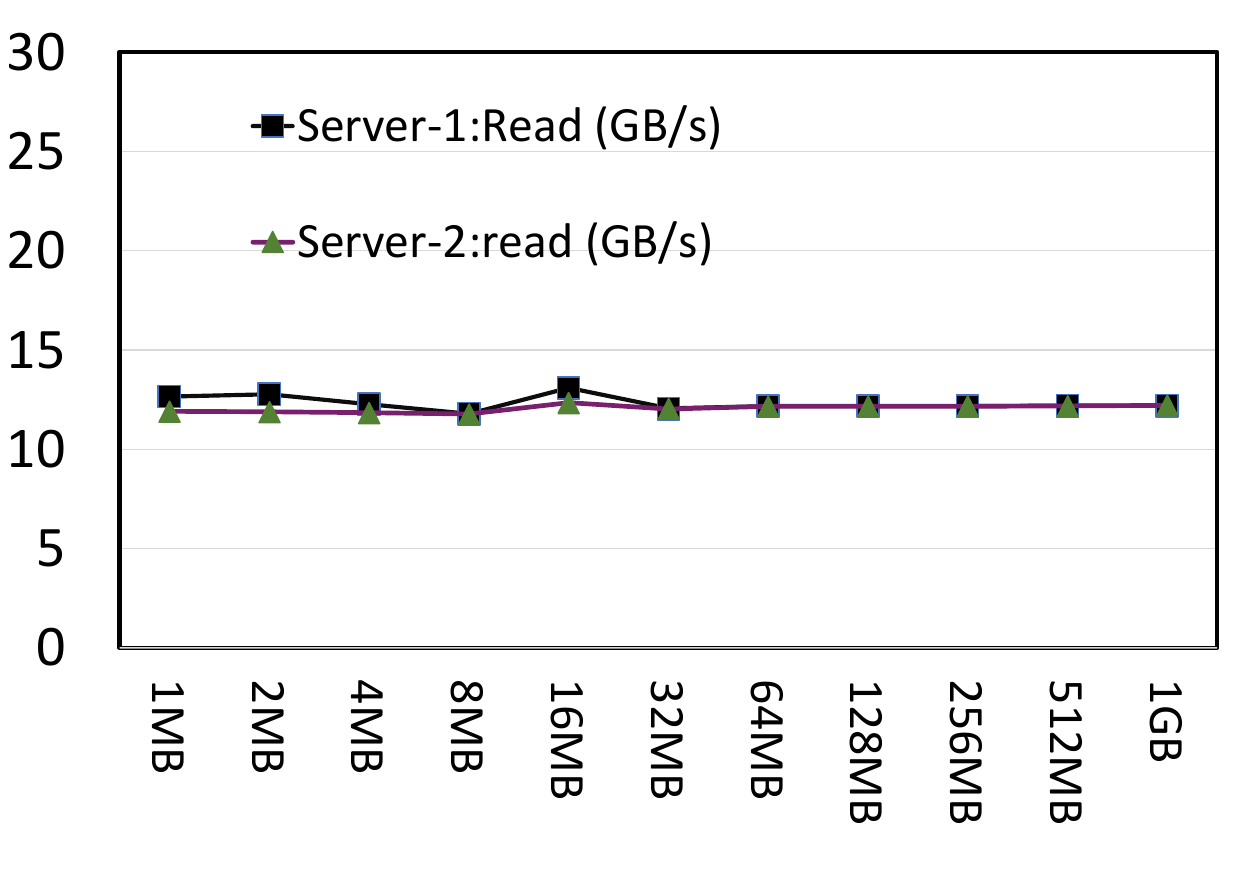}
    \caption{Concurrent GPU reads from the memory pool.}
    \label{fig:concurent_read}
\end{subfigure}
\hfill
\begin{subfigure}[t]{0.32\textwidth}
    \centering
    \includegraphics[width=\linewidth]{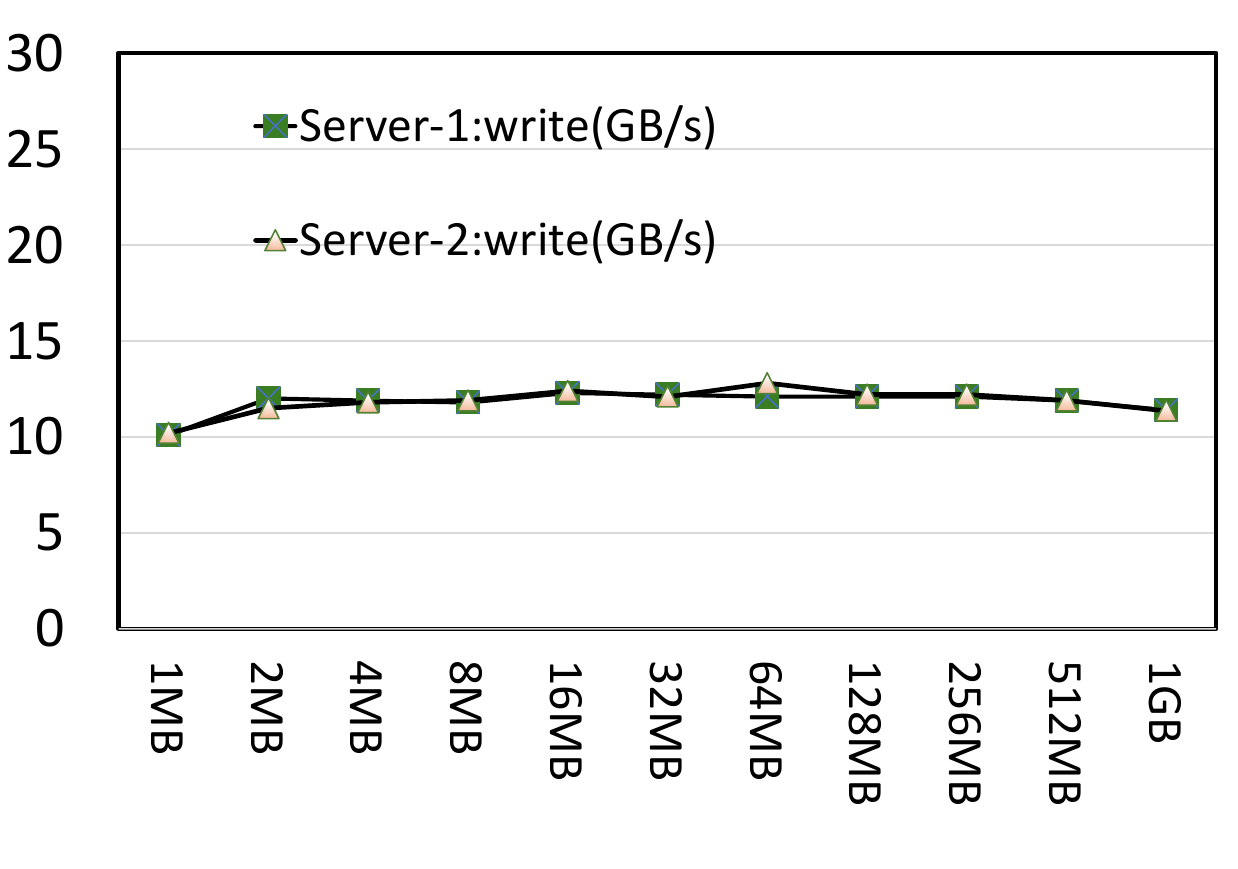}
    \caption{Concurrent GPU writes to the memory pool.}
    \label{fig:concurent_write}
\end{subfigure}

\caption{Performance characterization of the CXL shared memory pool. X-axis represents the transferred data volume. Y-axis represents the bandwidth(GB/s).}
\label{fig:gpu_cxl_2x2}
\end{figure*}
%To evaluate the performance of a shared CXL memory pool, this section presents microbenchmarks that measure both latency and bandwidth. The evaluation includes (1) CPU access latency to the CXL-backed address space, (2) GPU$\leftrightarrow$CXL data transfer bandwidth, (3) concurrent read and write streams targeting the same CXL device to study contention, and (4) concurrent streams targeting different CXL devices to study scaling with more devices. These results quantify the basic cost of using the shared pool and the throughput that can be achieved under single-stream and multi-stream workloads. 
%To assess the performance of a CXL shared memory pool, 
We present our evaluation using microbenchmarks that measure memory latency and bandwidth \textcolor{dong}{to drive the design of \name}. The evaluation covers four aspects: (1) CPU access latency to the CXL‑backed address space, (2) GPU$\leftrightarrow$CXL data transfer bandwidth, (3) concurrent read/write streams directed to the same CXL device to examine contention, and (4) concurrent streams targeting different CXL devices to evaluate scalability. %as more devices are added. 
Together, these results characterize the fundamental overheads of using the CXL shared memory pool and the achievable throughput under both single‑stream and multi‑stream workloads. \textcolor{fix}{Section~\ref{sec:evaluation_setup} shows the hardware setup for our test.}

%\textcolor{fix}{\textbf{CPU access latency to the CXL-vacked address space.}} We use Intel Latency Checker (MLC) to test the access latency for the local DRAM. Table~\ref{tab:mlc_latency} show that the CXL memory pool has higher access latency than local DRAM. Local DRAM is 214,ns, while the CXL pool is 658,ns, which is about $3.1\times$ higher. This gap is expected because DRAM is attached directly to the CPU memory controller, while CXL memory is reached through the PCIe/CXL link and the CXL switch, which adds extra hop and protocol overhead. Even with this higher latency, the CXL pool can still be useful for capacity expansion and for data that is not on the critical per-instruction path, especially when access is mostly streaming or batched.

\begin{table}[t]
    \centering
    \caption{MLC results.}
    \vspace{-10pt}
    \begin{tabular}{|c|c|c|}
    \hline
       & Local DRAM  & CXL memory pool \\ \hline
     Latency  & 214ns & 658ns \\ \hline
    \end{tabular}
    \label{tab:mlc_latency}
    \vspace{-10pt}
\end{table}

% \begin{figure*}[!t]
% \centering

% \begin{subfigure}[!ht]{0.33\textwidth}
%     \centering
%     \includegraphics[width=\linewidth]{ICS25-CXL/figures/BW.pdf}
%     \caption{Data transfer bandwidth, evaluated with single node-exclusive accesses.}
%     \label{fig:gpucxlbw}
% \end{subfigure}

% \begin{subfigure}[t]{0.33\textwidth}
%     \centering
%     \includegraphics[width=\linewidth]{ICS25-CXL/figures/concurrent_read.pdf}
%     \caption{Concurrent GPU reads from the memory pool.}
%     \label{fig:concurent_read}
% \end{subfigure}

% \begin{subfigure}[t]{0.33\textwidth}
%     \centering
%     \includegraphics[width=\linewidth]{ICS25-CXL/figures/Write+write.pdf}
%     \caption{Concurrent GPU writes to the memory pool.}
%     \label{fig:concurent_write}
% \end{subfigure}
% % \begin{subfigure}[t]{0.45\textwidth}
% %     \centering
% %     \includegraphics[width=\linewidth]{ICS25-CXL/figures/w+r.pdf}
% %     \caption{Concurrent GPU write and read accesses to the  memory pool. }
% %     \label{fig:concurent_rw}
% % \end{subfigure}

% \caption{Performance characterization of the CXL shared memory pool.}
% \label{fig:gpu_cxl_2x2}
% \end{figure*}

\textcolor{checked}{\textbf{CPU access latency to the CXL shared memory pool.}} We use Intel Latency Checker (MLC) to measure the access latency of local DRAM (\textcolor{dong}{local NUMA node}). As shown in Table~\ref{tab:mlc_latency}, %the CXL memory pool exhibits noticeably higher latency than local DRAM: 
the CXL memory pool shows $3.1\times$ higher latency than local DRAM. 
%Local DRAM access measures 214 ns, whereas the CXL pool reaches 658 ns--approximately $3.1\times$  higher. 
This difference is expected: DRAM is directly attached to the CPU’s memory controller, while the CXL memory is accessed through the PCIe/CXL link and CXL switch, introducing additional hops and protocol overhead. Despite the higher latency, the CXL pool remains valuable for capacity expansion and for data that do not lie on the critical per‑instruction path, particularly when access patterns are streaming or batched. 

%\textbf{GPU$\leftrightarrow$CXL data transfer bandwidth.}  To measure GPU$\leftrightarrow$CXL data transfer bandwidth, the CXL memory pool is first mapped into each server’s user space through DevDAX (e.g., \texttt{/dev/dax*}). A fixed-size region is then pinned with \texttt{cudaHostRegister}, so it can be used as a CUDA host buffer for DMA transfers. Bandwidth is measured using timed, repeated \texttt{cudaMemcpyAsync} operations between a GPU buffer and the pinned CXL-mapped region, with warmup iterations to remove one-time overheads. Both directions are tested: GPU$\rightarrow$CXL (write into the pool) and CXL$\rightarrow$GPU (read from the pool). The transfer size is swept from small to large (e.g., MB to GB) to capture both latency-dominated and bandwidth-dominated regimes, and the achieved bandwidth is reported as $\text{bytes transferred} / \text{time}$.

\textbf{GPU$\leftrightarrow$CXL data transfer bandwidth.}  To measure GPU$\leftrightarrow$CXL data transfer bandwidth, the CXL memory pool is first mapped into each server’s user space through DevDAX (e.g., \texttt{/dev/dax*}). A fixed‑size region is then pinned using \texttt{cudaHostRegister}, allowing it to serve as a CUDA host buffer for DMA transfers. The bandwidth is evaluated using timed, repeated \texttt{cudaMemcpyAsync} operations between a GPU buffer and the pinned CXL‑mapped region, with warmup iterations included to eliminate one‑time initialization overheads. Both transfer directions are measured—GPU$\rightarrow$CXL (writes to the pool) and CXL$\rightarrow$GPU (reads from the pool). The transfer size is swept from small to large (e.g., MB to GB) to capture both latency‑dominated and bandwidth‑dominated behaviors, and the resulting bandwidth is reported as ``$\mathrm{bytes\  transferred}/\mathrm{time}$''.

%The evaluation proceeds in two steps. First, a single-server test measures the peak bandwidth when only one server performs transfers to/from the shared CXL pool, which gives the best-case per-server bandwidth without contention. Second, a two-server test runs the same benchmark concurrently on two servers, where both servers perform GPU$\rightarrow$CXL writes, CXL$\rightarrow$GPU reads, or one reads while the other writes. This setup quantifies how the shared pool and switch links behave under cross-server contention, and whether bandwidth scales, saturates, or drops when multiple servers access the pool at the same time.

The evaluation proceeds in two stages. First, a single‑server test measures peak bandwidth when only one server transfers data to or from the CXL shared pool, providing a best‑case estimation of per‑server bandwidth without contention. Second, a two‑server test runs the same benchmark concurrently on both servers, with configurations where both issue GPU$\rightarrow$CXL writes, both issue CXL$\rightarrow$GPU reads, or one reads while the other writes. This setup reveals how the shared pool and switch fabric behave under cross‑server contention and whether bandwidth scales, saturates, or degrades with concurrent accesses from multiple servers. %%access the pool simultaneously. 

%\textbf{Results analysis.} 
Figure~\ref{fig:gpucxlbw} presents the measured bandwidth with exclusive accesses to the memory-pool devices. We observe that: (1) although the GPU is connected to the server via a PCIe Gen5 x16 link, the maximum sustained bandwidth for excessive read and write operations is constrained by the performance of the CXL memory module, which is attached through a PCIe Gen5 x8 interface. For relatively large message sizes (e.g., 1~MB), the achievable bandwidth approaches approximately 20~GB/s. We additionally conduct an experiment that issues multiple read or write requests to multiple CXL memory devices via multiple concurrent streams. Nevertheless, the aggregate peak bandwidth of all these requests does not exceed the peak bandwidth reported in Figure~\ref{fig:gpucxlbw}. We found that the primary cause is a hardware limitation of GPU, which is equipped with only a single DMA engine per data-transfer direction. Consequently, even in the presence of multiple CXL devices, a single GPU cannot fully saturate its PCIe link.

\textbf{Observation 1}: Bandwidth increases with message size, reaching ~20GB/s for 1MB transfers. However, performance is limited by a single CXL device on PCIe Gen5 x8 and the GPU’s single DMA engine per direction, which prevents full utilization of the PCIe Gen5 x16 link, even with multiple CXL devices.

Figure~\ref{fig:concurent_read} and Figure~\ref{fig:concurent_write} illustrate concurrent read and write operations initiated from multiple  servers. Results for message sizes below 1 MB are omitted because, during the evaluation, the data transfer time for smaller messages was so short that it was not possible to reliably initiate read and write operations concurrently. From the results, we observe that when multiple requests are issued to the same CXL device, the available bandwidth is distributed approximately evenly among the requests. This behavior is likely attributable to the scheduling or arbitration policy implemented by the CXL device’s internal controller. 

\textbf{Observation 2:} Issuing concurrent, similar requests to the same CXL device can substantially degrade the effective performance perceived by each individual server.

%\section{\name: A CXL-based collective communication for GPUs}

\section{Design of \name}

\subsection{Overview}
%There are two advantages of using CXL shared memory for collective GPU communications.

%Figure~\ref{fig:ncclpipeline} illustrates the traditional copy-RDMA pipeline for a send-received operation between two GPUs. This send-receive operation is the basic communication unit for forming other collective communication algorithms (such as Ring). This pipeline presents limitations. First, this method consumes GPU computing resources (SMs) and HBM bandwidth for copy operations between the user buffer and the FIFO buffer. This causes resource contention with concurrent computation and lead to performance degradation, forcing users to trade off resources between communication and computation.Second, this approach requires data to be segmented and transferred through multiple RDMA requests to establish the copy-RDMA pipeline. This process necessitates GPU-CPU synchronization at each pipeline stage and, more critically, limits each RDMA operation to a single data chunk. Finally, each pipeline is managed by a dedicated thread block, known as a “Channel.” This static binding between pipelines and thread blocks requires allocating separate FIFO buffers for each independent pipeline. As a result, increasing the number of thread blocks to accelerate a collective operation leads to higher GPU memory consumption.

\begin{figure}[t]
    \centering
    \includegraphics[width=1\linewidth]{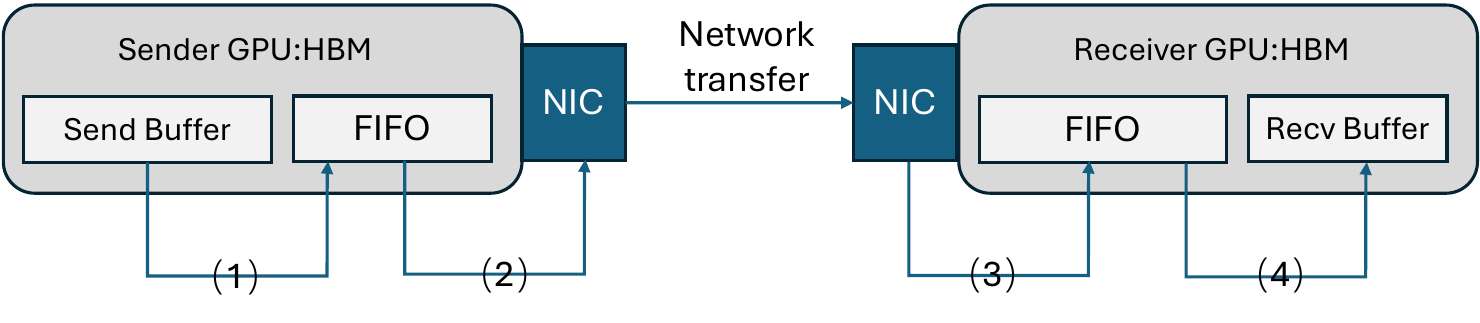}
    \caption{The traditional copy-RDMA communication pipeline in NCCL.}
    \label{fig:ncclpipeline}
\end{figure}

%%%\textbf{Fully CPU-driven programming model.}
The traditional GPU collective communication has limitations. Figure~\ref{fig:ncclpipeline} illustrates the traditional copy–RDMA pipeline used for send or receive operations between two GPUs. 
This send–receive primitive serves as the foundational communication unit for building higher‑level collective communication algorithms such as the ring-based one~\cite{woolley2015nccl}. \textcolor{camera}{The CPU will be responsible for launching the GPU kernel (including the whole send–receive pipeline) for each rank.}  This pipeline introduces several limitations. First, it consumes GPU compute resources (streaming multiprocessors) and HBM bandwidth to copy data between the user buffer and the FIFO buffer \textcolor{dong}{associated with the pipeline}. These copy operations contend with concurrent GPU computation, degrading performance and forcing the users to balance resources between communication and compute. Second, the pipeline requires data to be partitioned and transferred through multiple RDMA requests, each forming a stage of the copy–RDMA pipeline. Pipelining necessitates frequent GPU–CPU synchronization, as the CPU must intervene to verify kernel completion before dispatching the subsequent RDMA request. This introduces a control-plane overhead at every stage, effectively serializing the task-handover process. More critically, this restricts each RDMA operation to a single data chunk. Finally, each pipeline is managed by a dedicated thread block in GPU, or ``Channel''. This static binding requires allocating a separate FIFO buffer for every independent pipeline. As a result, increasing the number of thread blocks to accelerate a collective operation proportionally increases GPU memory consumption.

%With CXL shared memory pool, the data access interfaces are analogous to those used for local DRAM. Consequently, developers are freed from managing the low-level network stack, complex work request preparations, and performance optimizations required by RDMA-based communication. Furthermore, the elimination of cross-component synchronization overhead for GPUs simplifies the coordination between data transfers and computation. The pseudocode in Listing~\ref{lst:pseudo_programming_model} illustrates the core structure of a primitive task in the proposed programming model for collective communication. The task assumes that a shared memory pool has already been mapped and registered in the CUDA address space, allowing direct memory transfers between host and device. The task begins by writing data from GPU memory to the memory pool using \texttt{cudaMemcpy} with the \texttt{cudaMemcpyDeviceToHost} flag. After the data is written, a synchronization step ensures that all participating tasks have completed their write operations before proceeding. If further operations are required, such as reduction or aggregation, data is read back from the memory pool into GPU memory using \texttt{cudaMemcpyHostToDevice}. This model enables flexible and modular support for collective operations by decoupling communication and computation, while maintaining explicit control over memory transfers.

With a CXL shared memory pool, data access becomes analogous to accessing local DRAM. As a result, developers are relieved from managing low‑level network protocols, complex work‑request preparation, and performance tuning typically required for RDMA based communication. Moreover, eliminating \textcolor{dong}{CPU-GPU synchronization} simplifies coordination between data movement and computation.

Listing~\ref{lst:pseudo_programming_model} presents the core structure of a communication primitive. The shared memory pool is mapped and registered in the CUDA address space, enabling direct memory transfers between the node and CXL device. Execution begins by writing data from GPU memory to the memory pool using \texttt{cudaMemcpy} with the flag \texttt{cudaMemcpyDeviceToHost}. After the write completes, a synchronization ensures that all participating tasks have finished their data transfers before proceeding. If additional operations—such as reduction or aggregation—are required, data is then read back from the memory pool into GPU memory using \texttt{cudaMemcpyHostToDevice}. This method provides flexible and modular support for collective operations by decoupling communication from computation while maintaining explicit control over memory transfers.

%new programming model
\begin{lstlisting}[float,caption={Core structure of a collective communication primitive in \name.}, label={lst:pseudo_programming_model}]
void primitive_collective_comm_task(...){

    // memory pool is already mapped and registered into CUDA space.
    ...
    // write the data to the pool
    cudaMemcpy(dst, src,size, cudaMemcpyDeviceToHost); 
    
    synchronization ;
    
    // do the read and reduce operation if needed
    cudaMemcpy(dst, src,size, cudaMemcpyHostToDevice);

    // do the reduction operation if needed
}
\end{lstlisting}

\begin{figure*}
    \centering
    \includegraphics[width=0.9\linewidth]{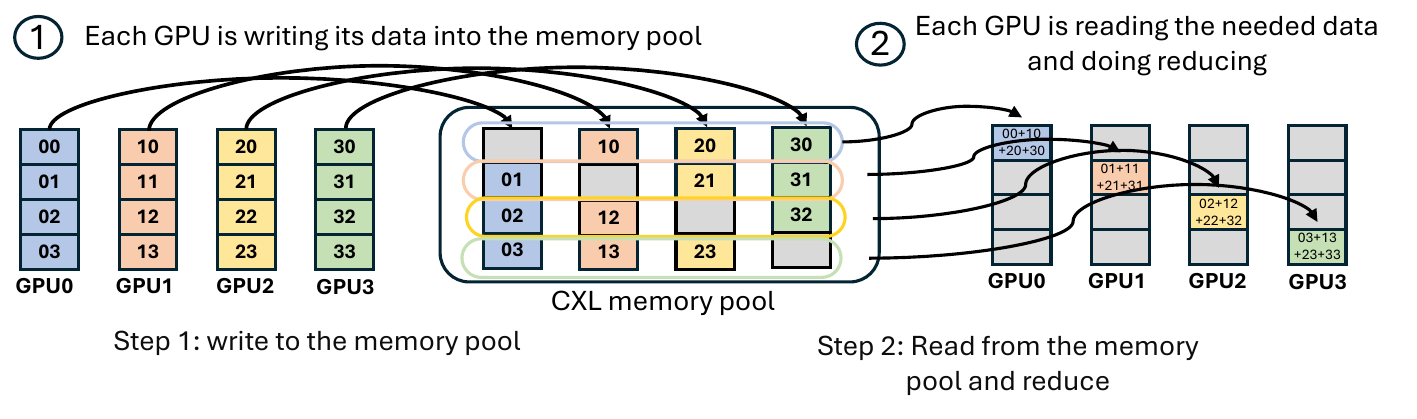}
    \caption{An example: ReduceScatter with four GPUs via a CXL shared memory pool.}
    \label{fig:reduce_scatter_cxl}
    \vspace{-5pt}
\end{figure*}

%\textbf{Memory management.} In \name, each server will have a unified address space with simplified space control. During startup, the BIOS on each host detects the attached CXL devices and reserves a contiguous physical address space for them. \name then leverages this foundation by having each host manage the CXL memory pool in Direct Access (DAX) mode. In this mode, the CXL memory is exposed as a block device, allowing user-space processes to use the mmap() to map the entire memory region directly into their virtual address spaces. This direct-mapping mechanism is the key to both resource partitioning and data sharing.  Crucially, this approach provides a unified memory view across all servers, facilitating simple and efficient memory management.

In \name, each node has a unified address space with simplified space control. During startup, the BIOS on each node detects the attached CXL devices and reserves a contiguous physical address space for them. \name then leverages this foundation by having each node manage the CXL memory pool in DAX mode. In this mode, the CXL memory is exposed as a block device, allowing user-space processes to use \texttt{mmap()} to map the entire memory region directly into their virtual address spaces. This direct-mapping mechanism is the key to both resource partitioning and data sharing. Crucially, this approach provides a unified memory view across all nodes, facilitating simple and efficient memory management.

\subsection{Challenges}
%Even the shared CXL memory pool provides  easiness, however, there are three main challenges. 
The design of \name faces three challenges. 

\textbf{1. Support for fine-grained interleaving at the cache-line level is not available.} The CXL shared memory pool can scale capacity by stacking multiple devices, but the memory pool lacks the hardware‑managed, cache‑line–level interleaving found in conventional multi‑channel DRAM systems, \textcolor{dong}{which leads to load imbalance across devices and limits memory bandwidth.} In the traditional DRAM, fine‑grained interleaving automatically stripes adjacent cache lines across memory channels, balancing requests and maximizing parallel bandwidth. In contrast, a CXL memory pool exposes each device as an independent target, so data placement is not automatically distributed across devices. Without an explicit placement mechanism, concurrent accesses from nodes or ranks may converge to the same device, creating hot spots, severe contention, and highly variable latency. As a result, the system cannot fully exploit the memory pool’s aggregate bandwidth, and overall performance can degrade substantially. To address this limitation, we design a data placement scheme that proactively distributes data and accesses across devices, reducing contention and improving utilization of available memory-level parallelism.

%\textbf{3. Sequentially data writing and data retrieval.} In a shared-memory–pool–based collective, data preparation and data retrieval are inherently dependent due to correctness requirements. Specifically, before a rank can read and reduce a peer’s contribution, it must guarantee that the peer has already written the corresponding data chunk into the memory pool. This readiness constraint introduces a strict producer–consumer ordering: ranks first publish their data to the pool, and only then can other ranks consume it by reading the data back to local memory and performing the reduction. A straightforward approach is to perform synchronous write and read operations for each rank: each rank first publishes its data and then begins reading the required data from the shared memory pool. However, this scheme leads to inefficiencies for certain collective operations. For instance, in an all-to-all primitive, each rank must publish \(N\) bytes and subsequently read \(N\) bytes; enforcing strict synchrony in these operations increases the overall execution time. By partitioning the outgoing data into smaller segments and exploiting the doorbell mechanism, \name can overlap publication and retrieval, thereby performing these operations asynchronously.

\textbf{2. Sequential data publication and retrieval.} Using a shared memory pool for collective communication naturally introduces read-after-write (RAW) dependency: a rank must first publish their data to the pool, after which other ranks can safely retrieve it to perform the communication. 
%However, this approach becomes inefficient for certain collective operations. For example, in an all‑to‑all primitive, each rank must publish $N$ bytes and subsequently read $N$ bytes; enforcing strict synchrony serializes these steps and increases overall execution time. 
RAW enforces a strict order on memory operations and limits communication performance. To improve the performance, \name partitions communication data into smaller segments and leverages a doorbell mechanism to establish execution order. This method overlaps data publication and retrieval, enabling them to proceed asynchronously and improving performance.

\textbf{3. Difficulty in supporting read-after-write.} %In \name, each rank independently writes its local data into the shared memory region, and other ranks subsequently read the required data to complete the collective communication operation. On a single machine, read‑after‑write (RAW) correctness is naturally ensured by hardware cache coherence and atomic memory operations. However, the CXL memory hardware does not provide full cache coherence or atomicity across nodes or CXL devices. Consequently, a rank cannot reliably determine when another rank’s write has completed and become globally visible. Without a RAW guarantee, the consumer may observe stale or partially written data, making correct cross‑rank synchronization impossible. To address this limitation, an explicit synchronization and notification mechanism is required to signal data readiness before reads can safely proceed. We introduce a lightweight doorbell mechanism that allows the consumer to wait for specific data to become available, ensuring correctness with minimal overhead.
\textcolor{checked}{In \name, each rank independently writes its local data into the shared memory region, and other ranks subsequently read the required data to complete the collective communication operation. On a single node, RAW correctness is naturally respected by hardware cache coherence. However, the CXL hardware does not provide full cache coherence across nodes or CXL devices. Consequently, a rank cannot reliably determine when another rank’s write has completed and become globally visible. Without a RAW guarantee, cross-rank synchronization is difficult to establish, and the consumer may observe stale or partially written data. %making correct cross‑rank synchronization impossible. 
To address this limitation, an explicit synchronization and notification mechanism is required to signal data readiness before reads can safely proceed. We introduce a lightweight doorbell mechanism on the memory pool that allows the consumer to wait for specific data to become available, ensuring correctness with minimal overhead.}

\name has three major components, discussed as follows.

%\section{\name}
%\subsection{Overview}
%In \name, there are 3 key parts to enable efficient collective communication between GPU through CXL shared memory pool: 1) a doorbell mechanism to notify the other ranks that the data is ready to be accessed. 2) Data placement to utilize the aggregated bandwidth. 3) Fine-grained chunk-based data exchanging to exploit the dual-duplex nature of CXL link.
%\subsection{Lightweight lock mechanism:doorbell}

\subsection{Bandwidth Aggregation}
%Bandwidth Aggregation
%The CXL shared memory pool provides a scalable interface between host processors and memory devices. Both capacity and bandwidth can be increased by integrating additional CXL devices into the memory enclosure. This approach enhances the aggregate bandwidth but does not support fine-grained, cacheline-level interleaving across devices. Consequently, even though the total available bandwidth is high when multiple devices are deployed, suboptimal data placement can lead to concentrated, concurrent memory accesses to the same CXL device, resulting in underutilization of the overall memory pool. As illustrated in Figure~\ref{}, when two servers simultaneously access the same CXL device, each server effectively obtains only 50\% of that device’s total bandwidth.

%\textcolor{red}{(this paragraph can be significantly reduced if we want to save space.)} 
The CXL shared memory pool provides a scalable interface between host processors and memory devices. Both capacity and bandwidth can be expanded by integrating additional CXL devices into the memory enclosure. While this increases aggregate bandwidth, the pool does not support fine‑grained, cacheline‑level interleaving across devices as conventional multi‑channel DRAM does. %As a result, even with multiple devices installed, suboptimal data placement can cause concurrent accesses from different servers to converge on the same CXL device, leading to contention and underutilization of the overall memory pool. 
As illustrated in Figure~\ref{fig:concurent_read} and Figure~\ref{fig:concurent_write}, when two nodes simultaneously access the same device, those memory accesses converge on the same CXL device, and each node effectively receives only half of that device’s available bandwidth. To fully exploit the aggregated bandwidth, we introduce a software-level interleaving mechanism to enable data distribution across CXL devices. %the data of each rank across CXL devices.  

\textcolor{checked}{The conventional approach for interleaving often relies on the maintenance of a large pool of memory blocks, from which each rank dynamically acquires a block when it needs to write data~\cite{corbet2023cxl, buddy_slab_allocator}. However, this approach necessitates the management of substantial metadata and the support of parallel allocation, since all ranks may begin writing their data into the pool concurrently once a request is issued. Furthermore, it requires careful placement of allocations across different devices to minimize concurrent read/write conflicts. %as much as possible.
In \name, we propose an alternative mechanism: preallocated, model-guided memory regions for each rank. The key observation is that collective communications exhibit regular and predictable access patterns, which can be systematically characterized and therefore proactively arranged across CXL devices.} \textcolor{dong}{We discuss our model (or formula) to guide the data placement or interleaving across CXL devices as follows.}

\textbf{Variables.} We define the following  variables to illustrate our interleaving method. %better show how we do the interleaving. 
\begin{itemize}
    \item $ND$: the number of CXL memory devices in the pool.
    \item $DS$: the capacity of each CXL memory device.
    \item $rank\_id$: the identifier assigned to each process in a communicator. It is an integer from 0 to $N-1$, where $N$ is the total number of processes in the communicator. 
    \item $Data\_id$: the logical id of the data. This ID is a logical identifier used to refer to data in the $sendBuffer$. Its range depends on whether the data is sent to a single rank or distributed across multiple ranks. For example, in Figure~\ref{fig:reduce_scatter_cxl}, with 4 ranks, the $sendBuffer$ is divided into 4 segments, each destined for a different rank. In this case, data\_id ranges from 0 to 3.
    \item $DB_{Offset}$: the size of the preallocated doorbell buffer.
\end{itemize}

%In a collective communication, each rank  maintains a $sendBuffer$ and a $recvBuffer$. $sendBuffer$ stores the data to send to other ranks. $RecvBuffer$ is used to store the recieved data. As depicted in Figure~\ref{fig:reduce_scatter_cxl}, there two main steps for each rank to finish this collective communication request. For the first step, it will write the data from $sendBuffer$ to the memory pool. Then it will read the data from the memory pool to $recvBuffer$. During these two steps, each rank need to know where to send the data and also where to read the data.

In a collective communication primitive, each rank maintains a sendBuffer and a recvBuffer. %The $sendBuffer$ holds the data that the rank will send to others, while the $recvBuffer$ stores the data it receives. 
As illustrated in Figure \ref{fig:reduce_scatter_cxl}, each rank performs the collective in two main steps. First, it writes the data from its sendBuffer into the shared memory pool. Then, it reads the required data from the pool into its recvBuffer. Throughout these steps, each rank must know both where to write its outgoing data and where to read its incoming data.

\textbf{Determining the location to write the data.} As shown in Table~\ref{tab:nccl_primitives}, the collective communication primitives can be categorized into two types: (1) 1 to $N$ (or $N$ to 1), and (2) $N$ to $N$. For each category, we use a different interleaving method.

The following formulas determine how \name calculates the device location to store the data from the $sendBuffer$ for each rank for the type (1).  

\begin{equation}
\text{device\_index} = \text{data\_id} \% ND
\label{eq:device_index}
\end{equation}
% \begin{align}
% \text{device\_block\_id} &= \text{rank\_id} \times \left \frac{N_{\text{data}} + ND - 1}{ND} \right \notag \\
% &\quad + \left \frac{\text{data\_id} + \text{rank\_id}}{ND} \right
% \label{eq:device_block_id}
% \end{align}

\begin{equation}
\text{device\_block\_id} = \frac{\text{data\_id}}{\text{ND}}
\label{eq:device_block_id}
\end{equation}

\begin{align}
\text{device\_location} 
&= \text{DB\_offset} 
+ \text{device\_block\_id} \times \text{block\_size} \notag \\
&\quad + \text{device\_index} \times \text{DS}
\label{eq:device_loc}
\end{align}

Equation~\ref{eq:device_index} determines the target CXL memory device for a given data \textcolor{camera}{block}. %It consists of two main components. 
The term $(data\_id \bmod ND)$ indicates that \name employs a round-robin strategy to interleave data \textcolor{camera}{block} across all CXL  devices based on the identifier $data\_id$. This scheme evenly distributes data stored in sendBuffer for the root rank for the type (1) collectives.

\begin{figure*}[!ht]
    \centering
    \includegraphics[width=0.9\linewidth]{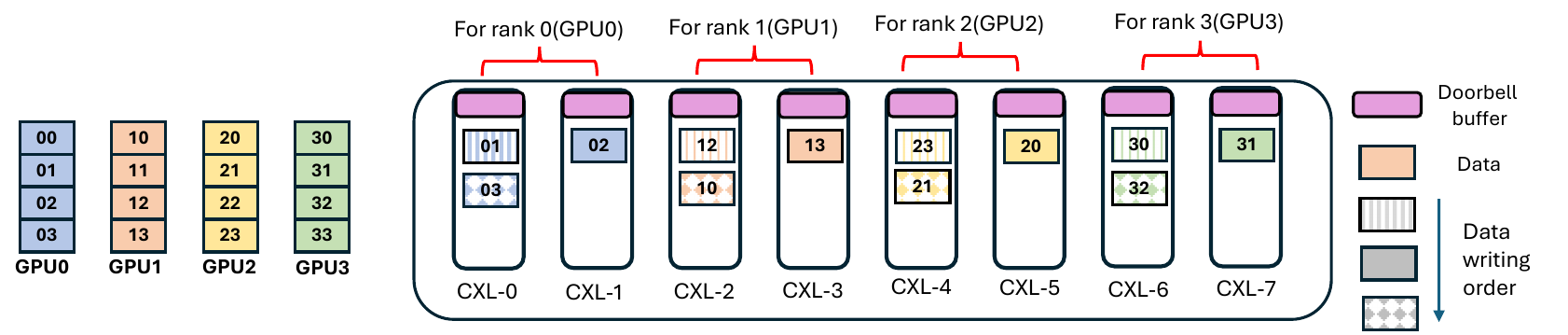}
    \vspace{-10pt}
    \caption{An example of spreading data across multiple CXL devices.}
    \label{fig:data_layout}
    \vspace{-10pt}
\end{figure*}

During data retrieval, each rank reads from the pool using a different offset. For instance, if the root rank is 0, rank 1 reads from \texttt{data\_0}, while rank 2 reads from \texttt{data\_1}. In this manner, the root rank publishes its data across all CXL devices in the memory pool, and other ranks access data from distinct devices in parallel. This design exploits the aggregated bandwidth of the memory pool and enhances parallel I/O efficiency.

Equation~\ref{eq:device_block_id} computes the identifier of a data block within a specific device. In this context, a data block denotes a logical block within the CXL memory space. The size of this block can be determined at runtime based on the message size associated with a given collective communication request. The resulting $device\_block\_id$ is further employed as an index to retrieve the corresponding doorbell entry from the doorbell buffer (discussed in Section~\ref{sec:locking}).

Equation~\ref{eq:device_loc} computes the final device location. It consists of three components: the preallocated doorbell buffer, the offset associated with the requested data blocks, and the offset corresponding to the $device\_index$ devices. By adding the mapped $baseAddr$, the resulting value specifies the address within the memory pool used for the write operation.

For the type (2) ($N$ to $N$), since all ranks write and read the data. The above method could incur concurrent read and write to the same devices. Hence, we propose another interleaving method that \textcolor{dong}{assigns different devices to different ranks}. The following formulas decide how we get the CXL memory location for a specific data.

\begin{equation}
\text{device\_id} = \text{data\_id} \% \text{device\_per\_rank}
\label{eqdevice_id}
\end{equation}

In Equation~\ref{eqdevice_id}, we introduce a  variable: $device\_per\_rank$, calculated by $ND/TOTAL\_RANK$. To calculate   $device\_block\_id$ and  $device\_location$, we use the same computation logic as in  Equations~\ref{eq:device_block_id} and \ref{eq:device_loc}. The only difference is that for each rank, the assignment is limited to a \textcolor{fix}{mutually exclusive} range of devices instead of \textcolor{fix}{all the devices}.

\textbf{Read data from the memory pool.} At this step, each rank computes the memory-pool address of the required data \textcolor{camera}{block} using Equations~\ref{eq:device_index}, ~\ref{eq:device_block_id} and ~\ref{eq:device_loc}. Each rank  then performs the reduction operation if the corresponding collective communication includes a reduction. At this step, different ranks still read data from different devices. 

\textbf{Example.} Figure~\ref{fig:data_layout} shows an example using \texttt{reduceScatter}. All ranks write their data according to the formulas described above. \textcolor{camera}{Each rank (GPU) will be assigned with two CXL devices. }
During the data publishing step, each rank writes exclusively to its own designated devices, thereby eliminating concurrent writes to the same device. 
For each rank, we establish a deterministic ordering of the data blocks to be published.  \name always publishes the data block \textcolor{camera}{starting from the $(rank\_id + 1)\%TOTAL\_RANKS$ and then goes to the next one. For instance, in Figure~\ref{fig:data_layout}, rank~0 first publishes \texttt{data-01} to \texttt{device-0}, followed by publishing \texttt{data-02} to \texttt{device-1}. All ranks start publishing their data simultaneously. When rank~0 reads \texttt{data-30} (waiting until the data becomes available) from \texttt{device-6}, rank~3 is simultaneously writing \texttt{data-31} to \texttt{device-7}.} In this manner, we avoid concurrent reads and writes targeting the same CXL device.

\begin{figure}[!t]
    \centering
    \includegraphics[width=0.9\linewidth]{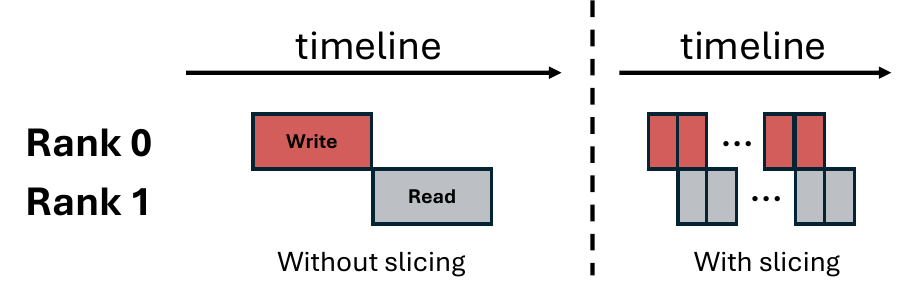}
    \caption{Communication overlapping.}
    \label{fig:slicing}
    \vspace{-20pt}
\end{figure}

\subsection{Asynchrony and Overlapping}
In Figure~\ref{fig:reduce_scatter_cxl}, a straightforward approach to preserve communication correctness is to introduce an explicit synchronization barrier between Step 1 and Step 2. This ensures that all data for communication are fully written to and visible in the memory pool before any subsequent accesses occur. However, this approach enforces strictly sequential read and write operations for each rank, thereby limiting performance.

\begin{comment}
We observe that the true data dependency arises only between a remote rank and the current rank: the remote rank must first write its data into the memory pool, and any rank that consumes this data must wait until the corresponding write has completed. Exploiting this observation, we allow each rank’s data publication (write) and data retrieval (read) to proceed asynchronously, as long as these inter-rank dependencies are respected.

Concretely, \name creates two distinct streams per rank: a `writeStream` for issuing write operations to the memory pool and a `readStream` for fetching data from the memory pool. By using these two streams, read and write operations at each rank can be executed in parallel without mutual blocking, while still preserving the required ordering constraints between producers and consumers.

Furthermore, \name partitions the entire data buffer into multiple data chunks. Each data chunk is associated with a corresponding doorbell. The rationale for this slicing strategy is that fine-grained data chunks enable overlap between the data publishing performed by the producer ranks and the data retrieval performed by the consumer (remote) ranks.
\end{comment}

We observe that true data dependencies arise only between a producing rank and the ranks that consume its data: the producer must first write its data into the memory pool, and any consumer must wait until that write has completed. Leveraging this observation, \name allows each rank’s data publication (write) and data retrieval (read) to proceed asynchronously, as long as these inter‑rank dependencies are respected.

In particular, \name creates two distinct streams per rank: a writeStream for issuing write operations to the memory pool and a readStream for fetching data from the pool. Using these two streams enables read and write operations at each rank to execute in parallel without blocking one another, while still preserving the necessary order constraints between producers and consumers. In addition, \name partitions the overall data buffer into multiple fine‑grained chunks, each associated with its own doorbell for synchronization. This chunking strategy allows producers to publish data incrementally and consumers to retrieve ready chunks immediately, thereby overlapping publication and retrieval. This method increases memory-level parallelism and reduces end‑to‑end communication latency. 

\begin{comment}
For example, consider a scenario in which rank-0 publishes \(N\) bytes of data into a shared memory pool. Without slicing, the remote rank must wait until all \(N\) bytes have been produced before it can begin transferring the data into its local buffers. During this period, the consumer rank remains blocked and repeatedly polls to determine whether the data is ready.

By contrast, when the data is partitioned into smaller chunks, the consumer rank can initiate data transfers as soon as the first chunk becomes available. Figure~\ref{fig:slicing} illustrates this slicing mechanism: the data publishing of rank-0 overlaps with the data retrieval of rank-1, thereby improving concurrency and reducing idle time.
\end{comment}

% \textcolor{red}{(the following paragraph can be simplified if we want to save space)}
Consider a case where rank 0 publishes $N$ bytes into the shared memory pool (see Figure~\ref{fig:slicing}). Without data chunking, the remote rank (rank 1) must wait for all $N$ bytes to be produced before starting \textcolor{camera}{to transfer data from the shared memory pool to its local buffers,}
% transfer of its local buffers, 
causing idling hardware and repeatedly polling the semaphore (the doorbell mechanism discussed in Section~\ref{sec:locking}). With partitioning into smaller chunks, the consumer can start transferring as soon as a chunk is available. As shown in Figure~\ref{fig:slicing}, this allows rank 0's publication to overlap with rank 1's retrieval, increasing concurrency and reducing idle time.

\subsection{Lightweight Locking Mechanism}
\label{sec:locking}
%Existing cross-server (or cross-node) locking mechanisms, such as centralized lock services~\cite{}, replicated lock services based on consensus~\cite{}, and lease-based locks~\cite{}, provide mutual exclusion for shared data. However, those mechanisms rely on cross-server messaging exposed in the critical path, which adds high latency and becomes a bottleneck for fine-grained synchronization. Recent work~\cite{} proposes two-level locking, but it is mainly designed for long-lived shared data where exclusive access must be maintained for a long time. However, we observe that, in the context of collective communication, data typically resides in the memory pool only briefly and is highly latency-sensitive. Maintaining heavyweight metadata and employing allocation operations to acquire locks further increases this latency. In \name, we therefore propose a lightweight, index-calculation-based mechanism that enables lock acquisition through a single, simple index computation.

\textcolor{dong}{To enable synchronization across nodes, we must establish a lock mechanism.} Existing inter‑node locking mechanisms—such as centralized lock services~\cite{center_lock, centrolizedlock1} and lease‑based locks~\cite{rodriguez2025distributedlockingperformanceanalysis}—provide mutual exclusion for shared data, but they rely on inter-node messaging in the critical path. This dependency introduces high latency and easily becomes a bottleneck for fine‑grained synchronization. Recent work~\cite{yoon2025tractdisaggregatedllmserving} proposes two‑level locking, but it is primarily designed for long‑lived shared data where exclusive access must be maintained for extended periods. In contrast, data involved in collective communication is short‑lived and highly latency‑sensitive. Maintaining heavyweight metadata or performing allocation operations to acquire locks  exacerbates this latency. To address the above challenges, \name introduces a lightweight, index‑calculation‑based mechanism that enables lock acquisition through a single, simple index computation.

%%Figure~\ref{fig:reduce_scatter_cxl} generally depicts the locking mechanism in \name. It has two steps. %using the memory pool for the collective communication has two phases. 
%In Step 1, each rank writes its output to the CXL shared memory pool. During this phase, only one writer updates a given data block, and all readers must wait until the write completes. In Step 2, ranks read the data block after it becomes ready. During this step, the data block is read-only and does not change. \textcolor{red}{(the above idea sounds very simple...)}

%As shown in Figure \ref{fig:reduce_scatter_cxl}, using the memory pool for collective communication proceeds in two phases. In phase 1, each rank writes its output into the shared CXL memory pool. During this stage, a given data block has exactly one writer, and all ranks that depend on this block must wait until the write completes. In phase 2, ranks read the data block once it becomes ready. At this point, the block is read‑only and remains unchanged throughout the remainder of the operation.

%To support this workflow, each data block has a dedicated semaphore stored in the shared memory pool, so that all servers can access it. \name assigns the permission to update the semaphore to the data owner (the rank that produces the data block). The semaphore has two states: $STALE$ and $READY$. Before the owner finishes writing the data block, the semaphore remains $STALE$. After the write completes, the owner sets it to $READY$.

\begin{figure*}[!ht]
    \centering
    \vspace{-10pt}
    \includegraphics[width=1\linewidth]{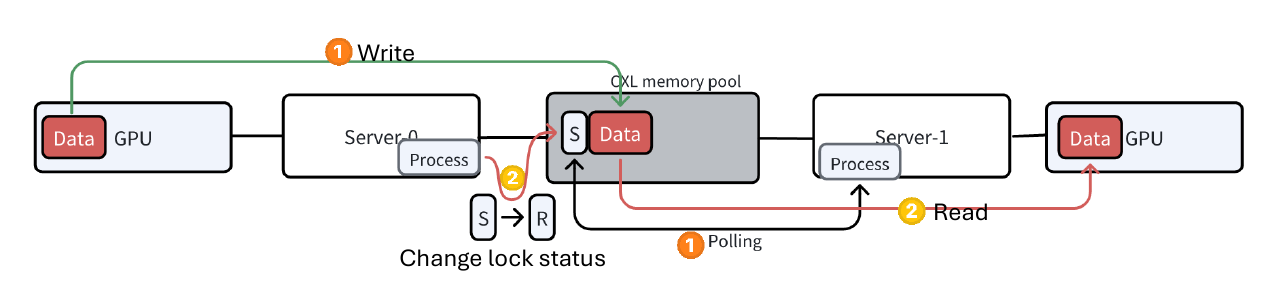}
    \vspace{-20pt}
    \caption{The workflow of locking mechanism in \name.}
    \label{fig:lock_workflow}
    \vspace{-15pt}
\end{figure*}

Figure~\ref{fig:lock_workflow} generally depicts the locking mechanism in \name with two steps. %It has two steps. In Step 1, each rank writes its output to the CXL shared memory pool. During this phase, only one writer updates a given data block, and all readers that need this block must wait until the write completes. In Step 2, ranks read the data block after it becomes ready. During this step, the data block is read-only and does not change.
Each data chunk is associated with a dedicated semaphore stored in the shared memory pool, making it accessible to all nodes. \name assigns update permission for this semaphore to the data owner—the rank responsible for producing the corresponding data chunk. The semaphore has two states, STALE and READY. Before the owner completes its write, the semaphore remains in the STALE state. Once the write finishes, the owner updates the semaphore to READY, signaling that the data chunk is safe for consumption. Figure~\ref{fig:lock_workflow} illustrates the above workflow.

Listing~\ref{lst:pseudo_code_rw} shows the pseudo code \textcolor{camera}{in CXL-CCL library}. In primitive\_task(...), the code uses a doorbell (\textcolor{dong}{semaphore})-based synchronization to establish the producer-consumer synchronization  between ranks when exchanging data through the memory pool. Lines 3–7 implement the publish (write) phase: the rank first copies the data from GPU memory into the pool using \texttt{cudaMemcpyDeviceToHost} (Line 4). After the copy completes, it selects the corresponding write doorbell entry (Line 5) and sets it to READY (Line 6) to announce that the data chunk is now valid and can be consumed. The doorbell update is then explicitly flushed (Line 7) to ensure visibility across sockets and prevent other ranks from observing stale cached values. Lines 8–13 implement the consumption (read) readiness check: the rank switches to the peer’s read doorbell (Line 9) and spins until the doorbell indicates the chunk is ready (Lines 10–13). If the doorbell remains in the STALE state, the consumer forces cache coherence by \textcolor{dong}{cache-line} flushing and re-reading (Line 11), ensuring it does not proceed with an outdated value. Here, this flushing operation only invalidates the doorbell in the cache. It does not change the status of the doorbell in the memory pool. 

Finally, once the doorbell confirms readiness, \textcolor{dong}{another rank} can safely read the corresponding data chunk from the pool and perform reduce operation (or other collective operations) if required (Lines 14–15), guaranteeing communication correctness while coordinating concurrent access among ranks.

\textbf{Pre-allocated doorbell Buffers.}  \name pre-allocates doorbell buffers in the CXL memory pool. %to avoid repeated allocations, which reduces the overhead of interacting with the CXL memory pool.
%to manage synchronization. Unlike existing approaches that dynamically allocate doorbell structures at runtime—often requiring multiple memory accesses—our design pre-allocates all necessary doorbell buffers during initialization. This avoids repeated allocations and significantly reduces the overhead of interacting with the CXL memory pool. Each collective communication request can directly compute the doorbell index based on a combination of \texttt{rank\_id} and \texttt{data\_chunk\_id}, allowing immediate access to the corresponding lock.
This design minimizes the number of memory transactions and avoids contention or latency spikes caused by dynamic memory management. As shown in Table~\ref{tab:mlc_latency}, accessing the CXL memory pool introduces 3.1× higher latency compared to local DRAM. Pre-allocating the doorbell buffers ensures that each rank can efficiently access synchronization primitives with minimal latency, leading to improved performance during collective communication.

\begin{samepage}
    
\begin{lstlisting}[float,caption={Pseudo code for how to use the doorbell-based synchronization across nodes in CXL-CCL.}, label={lst:pseudo_code_rw}]
void primitive_task(...){
    ...
    // write the data to the pool
    cudaMemcpy(dst, src,size, cudaMemcpyDeviceToHost);
    Doorbell* db_ptr = doorbell+index_write; // doorbell for the write
    *db_ptr = READY; // Make the doorbell ready
    flush_doorbell(db_ptr); // flush the change
    // read
    db_ptr = doorbell+index_read; // get the doorbell
    while(*db_ptr == STALE){
        flush_doorbell(db_ptr);
        sleep() ; // sleep for a short while
    }
    // do the read and reduce operation if needed
    ...
}
\end{lstlisting}
\end{samepage}

\section{Evaluation}
\label{sec:eval}

We implement \name on NCCL 2.28.3 with 1020 lines of code \textcolor{camera}{(including 30 modified lines and 980 added lines)}.

\subsection{Evaluation Setup} 
\label{sec:evaluation_setup}

\textbf{Hardware.} 
We use three nodes, each equipped with an Intel(R) Xeon(R) 6960P CPU (72 physical cores), 256\,GB of DRAM, and one NVIDIA H100 GPU with 80\,GB device memory, connected through a PCIe Gen5$\times$16 link. All three nodes connect to a TITAN-II CXL switch using PCIe/CXL Gen5$\times$16 links. Behind the switch, a memory box provides a CXL shared  memory pool using six Micron CZ120 CXL Type-3 memory cards, each with 128\,GB capacity and a PCIe/CXL Gen5$\times$8 interface, for a total pool size of 768\,GB. We disable the DDIO~\cite{yang2025beluga} to avoid using LLC for the data transfer and  set the memory pool address space as a non-cacheable area. \textcolor{dong}{Large-scale multi-host CXL shared memory pool platforms remain difficult to access commercially. Our evaluation platform, using  realistic hardware (not emulation or simulation),  represents the state of the art. \textcolor{camera}{\name is intended solely for research use and is not incorporated into Bytedance’s technologies.}}

\textbf{Software.} All nodes run Linux 6.2.6 with CUDA 12.8. 
%\name extends NCCL 2.28.3. 
We use nccl-tests 2.17.8 for evaluation.  $ncclSendRecv$ are not tested because they are point-to-point primitives.

%\textbf{Baselines.} Our baselines include one physical links: InfiniBand with 200Gb/s. All collective communication primitives are tested with variable message sizes—the message size ranges from 24B to 3GB. We categorize the primitives into two types: $N$ to $N$ and $N$ to 1 (or 1 to $N$). Table~\ref{tab:nccl_primitives} shows the primitives we test. The $ncclSend$ and $ncclRecv$ are not tested since these two are point-to-point primitives.

\textbf{Baseline.} \textcolor{dong}{We use InfiniBand with 200 Gb/s as the baseline}. All collective communication primitives are tested with variable message sizes, ranging from 1MB to 4GB. We categorize the primitives into two types: $N$ to $N$, and $N$ to 1 (or 1 to $N$). Table~\ref{tab:nccl_primitives} shows the primitives we test. 

\textcolor{dong}{We have three implementations of \name: (1) \nametwo, which is a full-featured implementation of \name; (2) \nameone has bandwidth aggregation but the aggregation occurs at a coarse-granularity(\textcolor{camera}{at data-block level)}. There is no ``asynchrony and overlapping''. (3) \name-Naive allocates memory in the memory pool sequentially without using memory interleaving. There is no ``asynchrony and overlapping''.}  

%\nametwo denotes the implementation that enables bandwidth aggregation. Furthermore, \nametwo also refers to the implementation that simultaneously provides bandwidth aggregation, asynchronous execution, and computation–communication overlap.

\begin{table*}
\centering
\caption{NCCL primitives. $N$ denotes the buffer size per rank, and \texttt{nranks} is the number of participating ranks.}
\small
\renewcommand{\arraystretch}{1.2}
\begin{tabular}{|l|c|c|p{7cm}|}
\hline
\textbf{Primitive} & \textbf{SendSize} & \textbf{RecvSize} & \textbf{Description} \\
\hline
\texttt{ncclAllReduce}     & $N$                   & $N$                    & All ranks contribute $N$ elements and receive reduction results. \\
\hline
\texttt{ncclBroadcast}     & $N$ (root only)       & $N$ (all ranks)        & Root rank broadcasts $N$ elements to all other ranks. \\
\hline
\texttt{ncclReduce}        & $N$                   & $N$ (root only)        & All ranks send $N$ elements to the root, which reduces and receives the result. \\
\hline
\texttt{ncclAllGather}     & $N$                   & $N \times \text{nranks}$ & Each rank gathers $N$ elements from every other rank. \\
\hline
\texttt{ncclReduceScatter} & $N$                   & $N / \text{nranks}$    & All ranks reduce $N$ elements; each rank receives a portion of $N / \text{nranks}$. \\
\hline
\texttt{ncclGather}        & $N$                   & $N \times \text{nranks}$ (root) & Each rank sends $N$ elements to the root which gathers all data. \\
\hline
\texttt{ncclScatter}       & $N \times \text{nranks}$ (root) & $N$           & Root scatters $N$ elements to other ranks from its buffer of the size $N \times \text{nranks}$. \\
\hline
\texttt{ncclAllToAll}      & $N$                   & $N$                    & Each rank sends $N/\text{nranks}$ elements to every other rank and receives the same from each. \\
\hline
\end{tabular}
\centering
\label{tab:nccl_primitives}
\end{table*}

\textbf{Scalability test.} We employ an emulation-based approach to conduct the scalability evaluation because of limited access to hardware. Specifically, we develop an emulator that reproduces system performance as the number of nodes is scaled. \textcolor{checked}{In this emulator, we assume that concurrent read or write requests targeting the same CXL device share the available bandwidth uniformly, aligned with the results presented in Section~\ref{sec:characteritics}. 
%For concurrent read–write mixtures, the bandwidth allocation and performance behavior follow the empirical results depicted in Figure~\ref{fig:concurent_rw}.
Requests directed to different CXL devices are assumed to be mutually independent, i.e., no performance interference is modeled across devices. The emulator is configured to employ a total of six CXL devices.}

\begin{figure*}[!ht]
    \centering
    % Row 1
    \begin{subfigure}{0.31\textwidth}
        \centering
        \includegraphics[width=\linewidth]{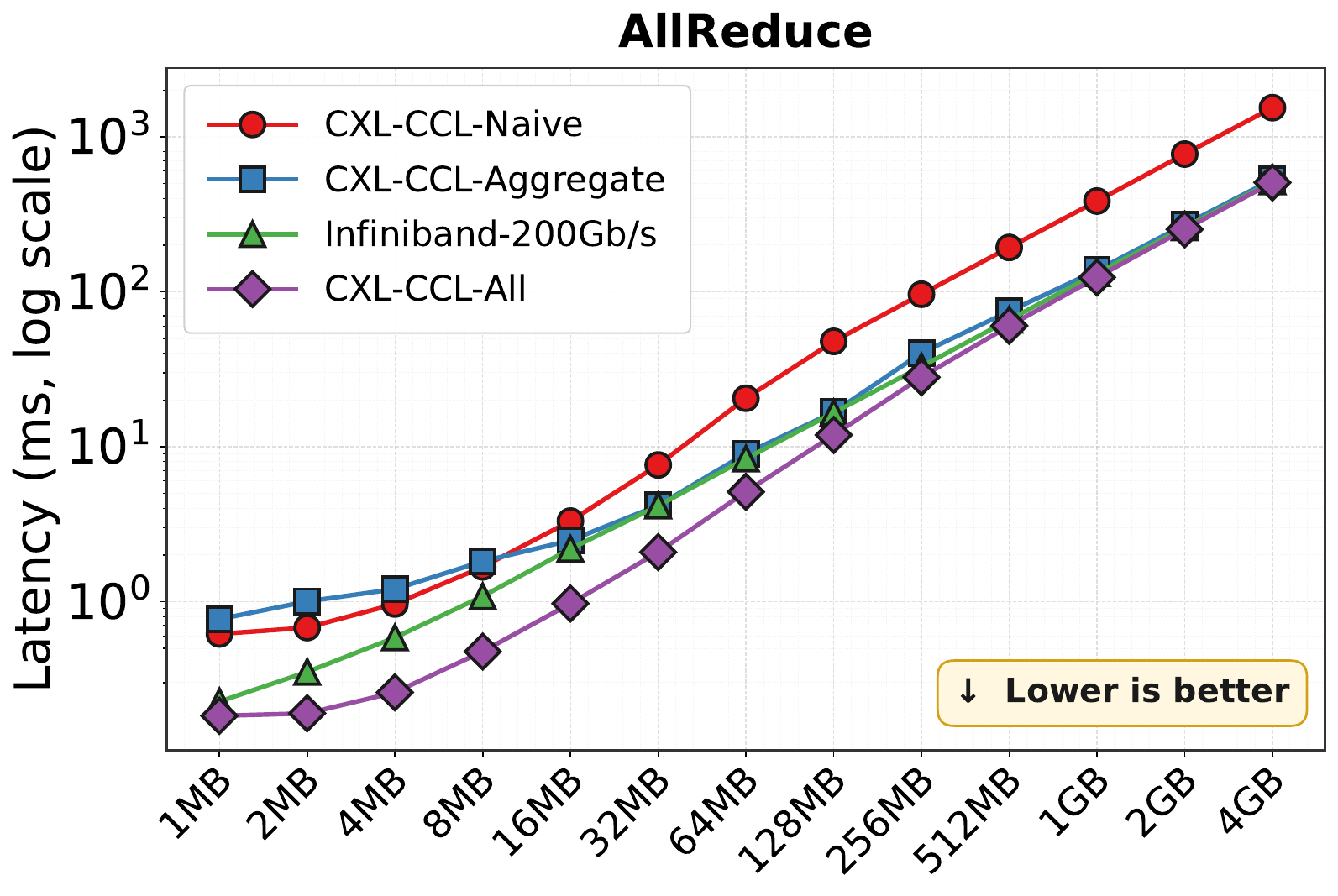}
        \caption{AllReduce.}
    \end{subfigure}
    \hspace{0.02\textwidth}
    \begin{subfigure}{0.31\textwidth}
        \centering
        \includegraphics[width=\linewidth]{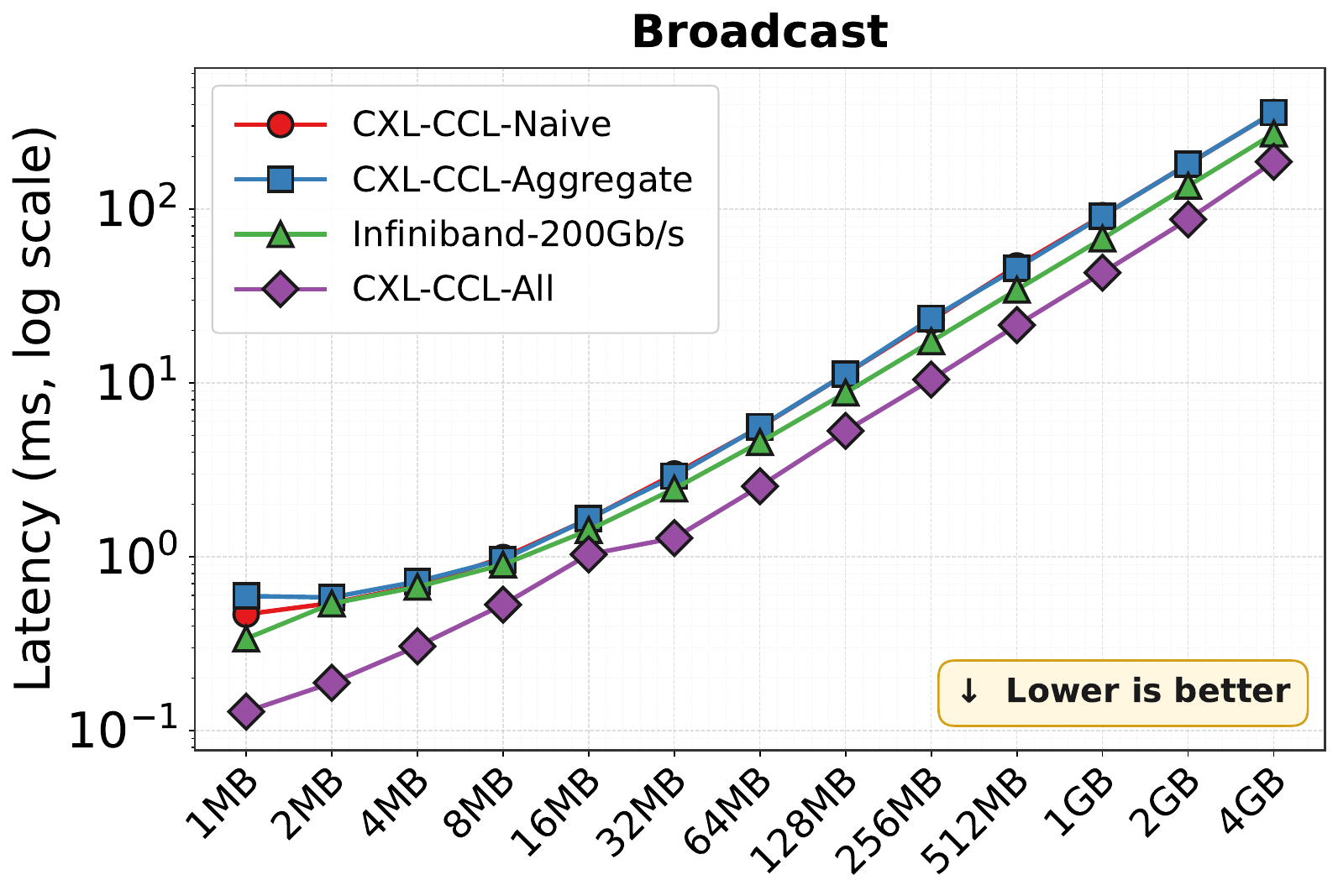}
        \caption{Broadcast.}
    \end{subfigure}
    \hspace{0.02\textwidth}
    \begin{subfigure}{0.31\textwidth}
        \centering
        \includegraphics[width=\linewidth]{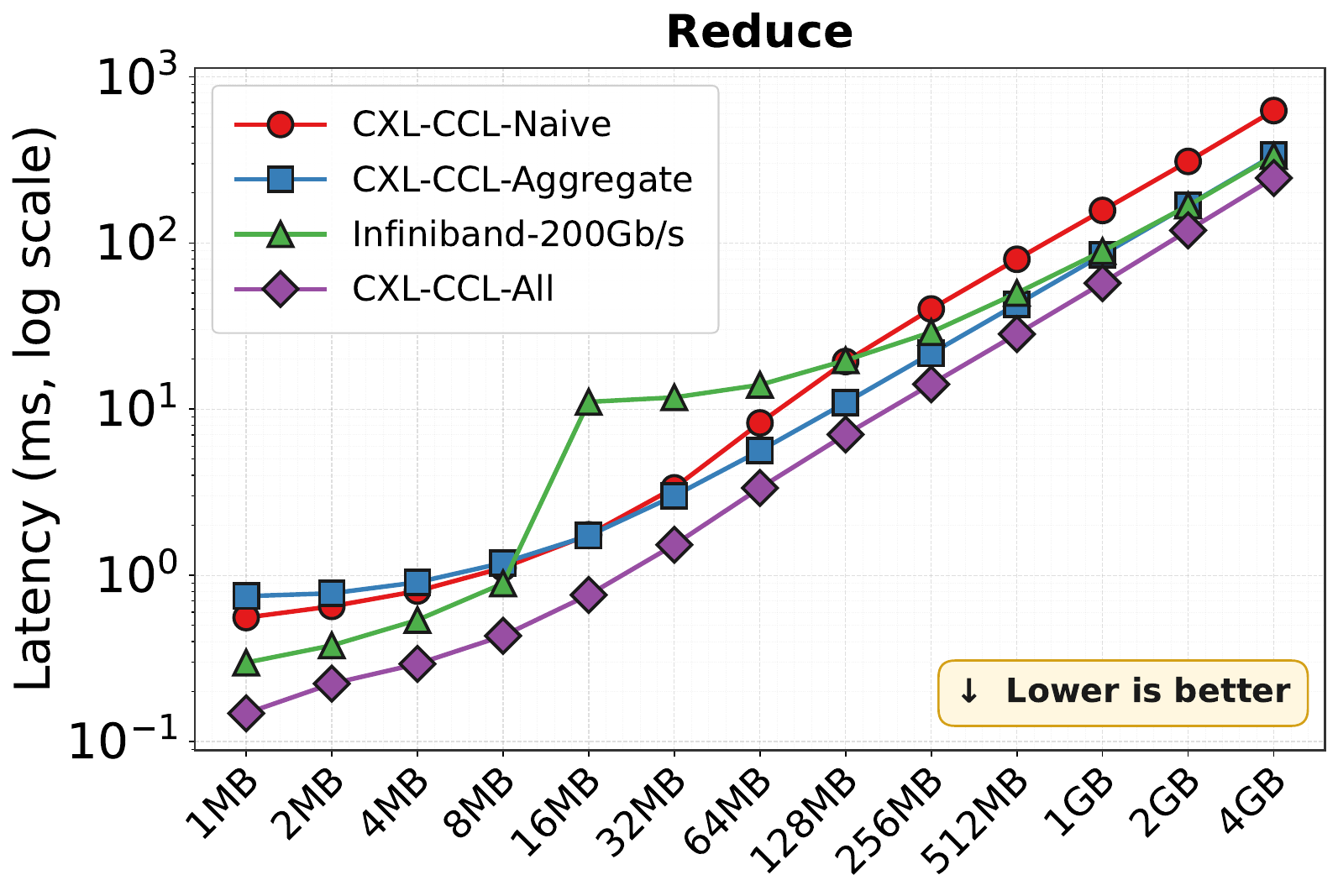}
        \caption{Reduce.}
    \end{subfigure}

    \vspace{1em}

    % Row 2
    \begin{subfigure}{0.31\textwidth}
        \centering
        \includegraphics[width=\linewidth]{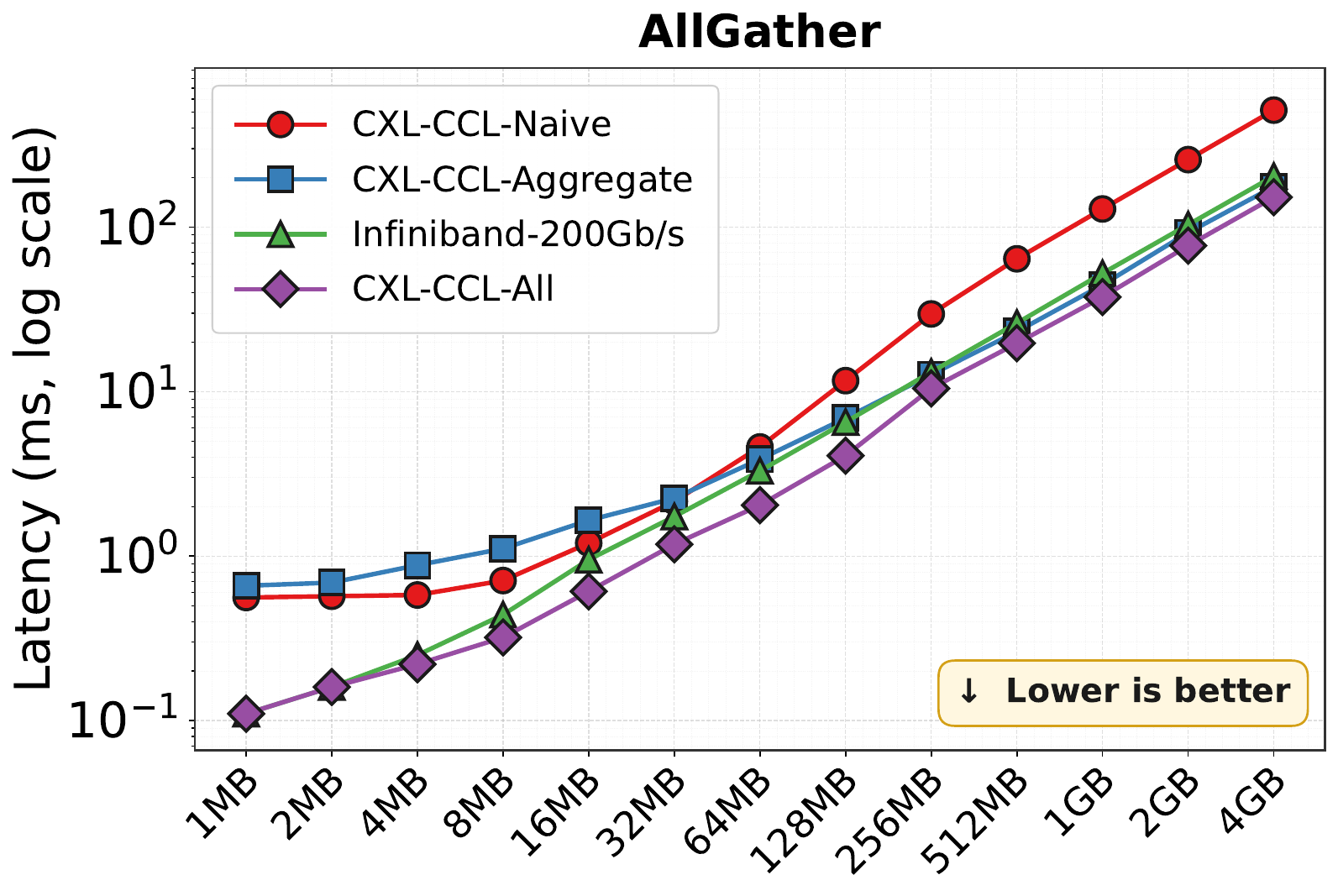}
        \caption{AllGather.}
    \end{subfigure}
    \hspace{0.02\textwidth}
    \begin{subfigure}{0.31\textwidth}
        \centering
        \includegraphics[width=\linewidth]{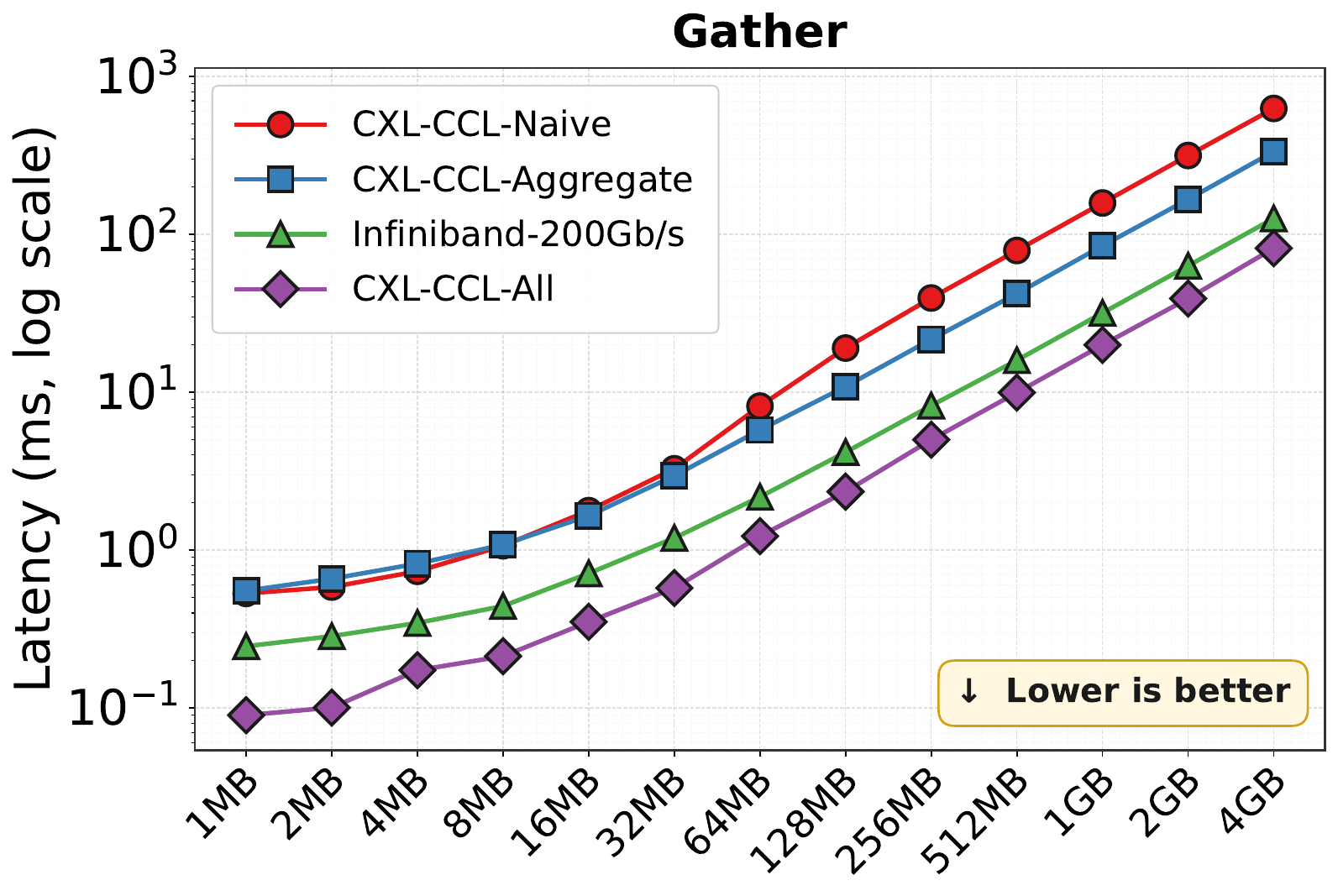}
        \caption{Gather.}
    \end{subfigure}
    \hspace{0.02\textwidth}
    \begin{subfigure}{0.31\textwidth}
        \centering
        \includegraphics[width=\linewidth]{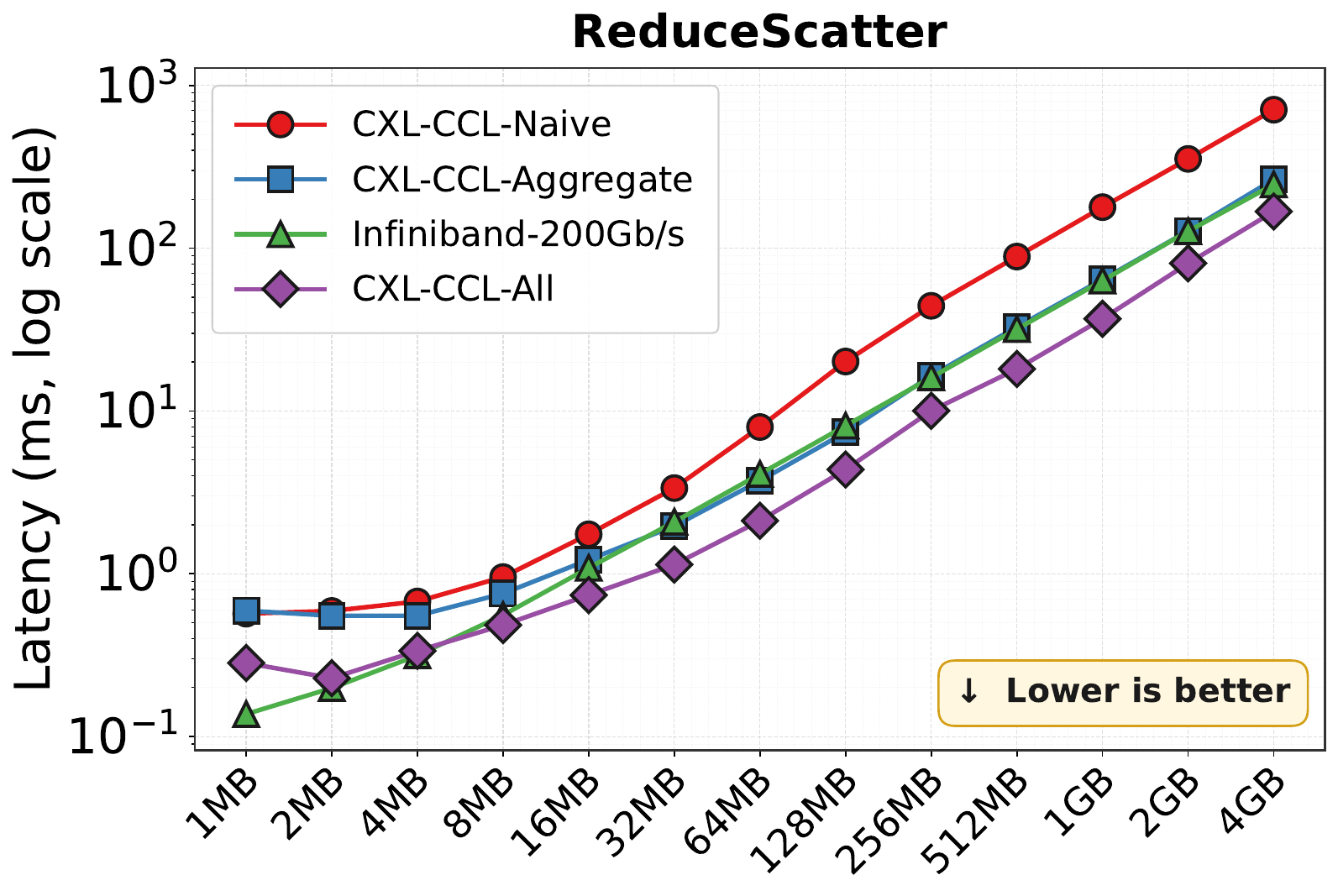}
        \caption{ReduceScatter.}
    \end{subfigure}
    \vspace{1em}

    % Row 3 (only 2 figures)
    \begin{subfigure}{0.31\textwidth}
        \centering
        \includegraphics[width=\linewidth]{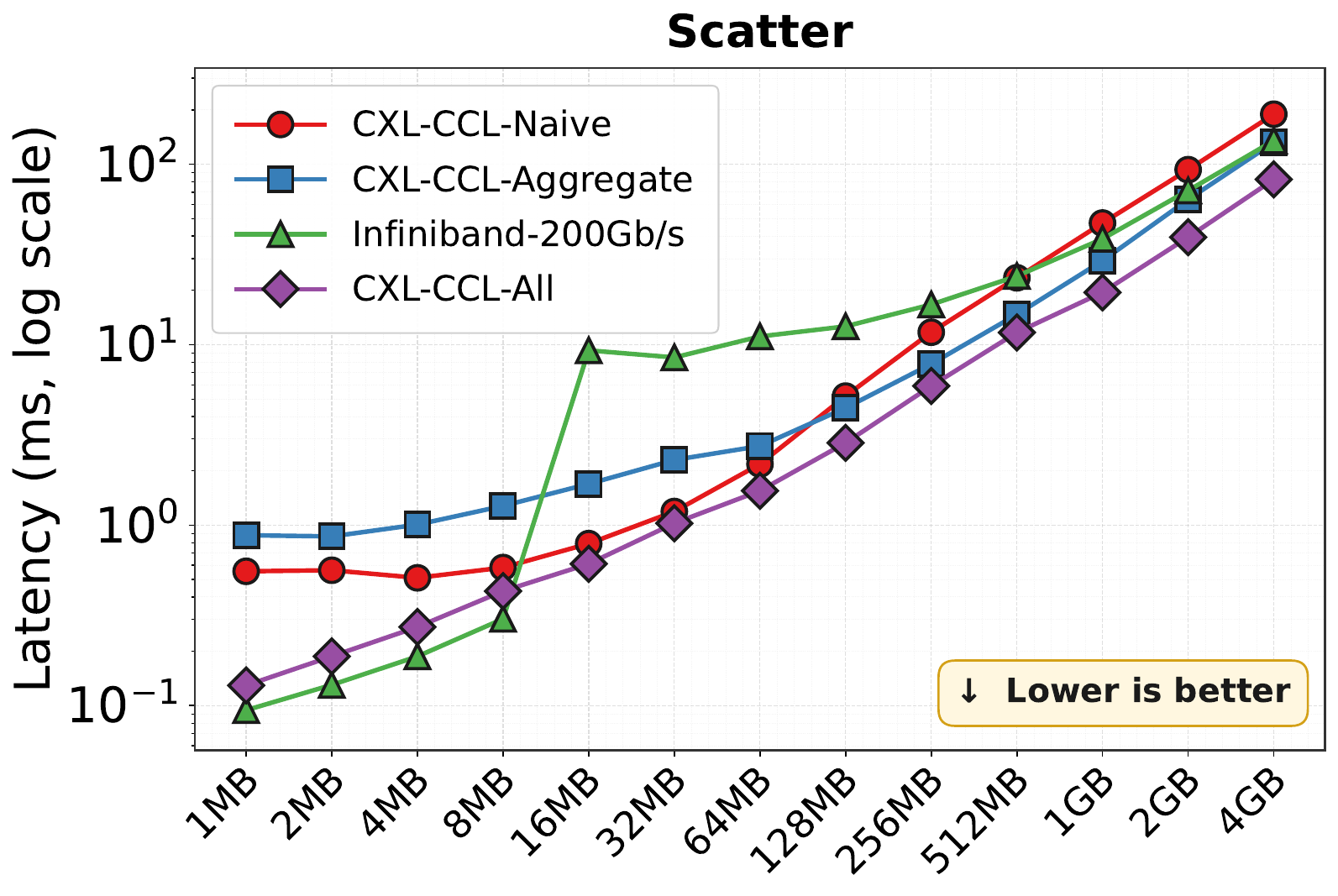}
        \caption{Scatter.}
    \end{subfigure}
    \hspace{0.02\textwidth}
    \begin{subfigure}{0.31\textwidth}
        \centering
        \includegraphics[width=\linewidth]{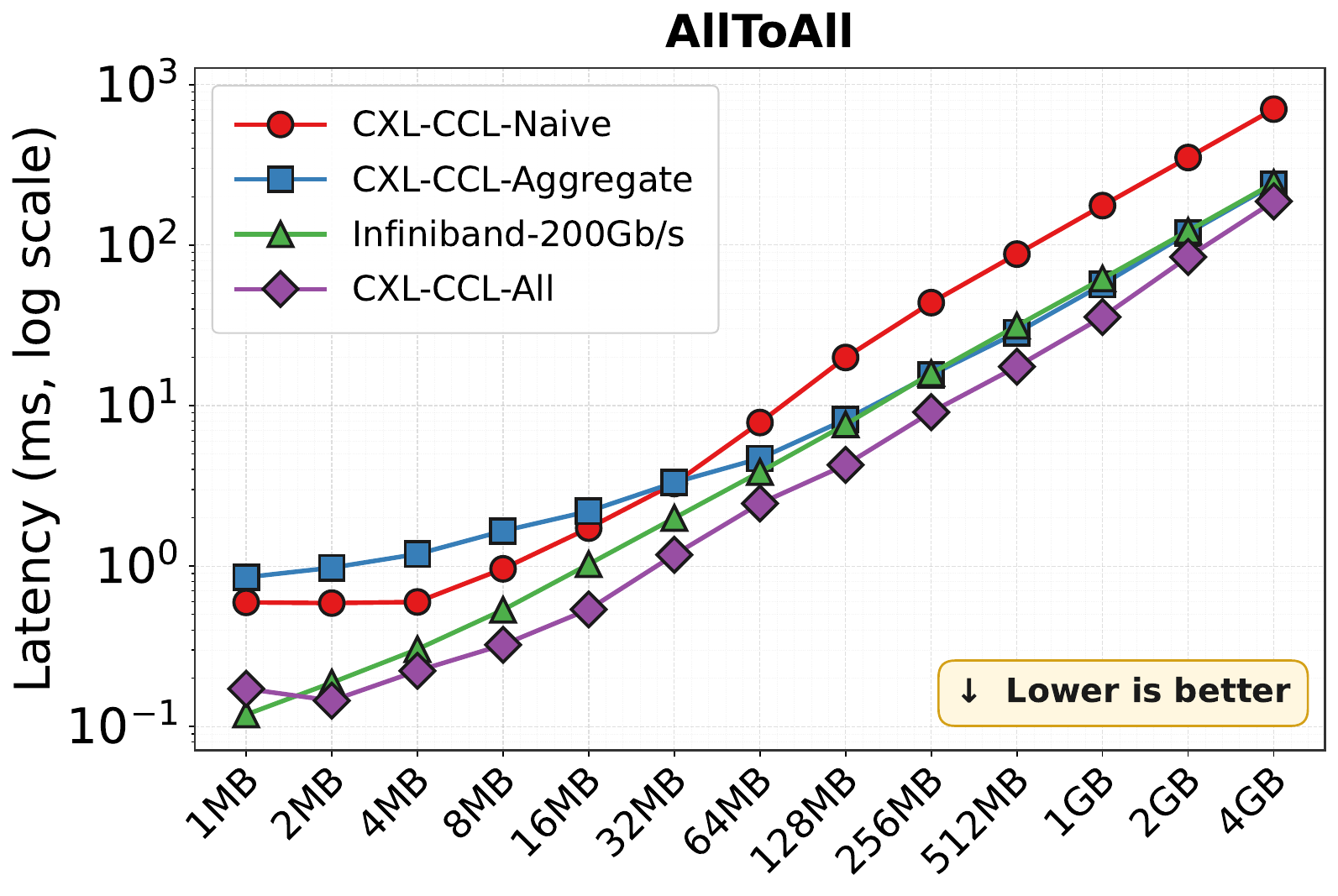}
        \caption{AllToAll.}
    \end{subfigure}
    \vspace{-10pt}
    \caption{\name performance using the CXL shared memory pool. \textcolor{dong}{The x axis is the message size.}}
    \label{fig:nccl-all}
    \vspace{-15pt}
\end{figure*}

\vspace{-10pt}
\subsection{Overall Performance}
%Overall performance.
Figure 9 reports the overall performance of CXL-CCL across NCCL primitives.
%shows the performance of major collective communication primitives. 

\textbf{AllReduce} implements an N-to‑N pattern in which each rank must independently reduce data from all other ranks. Because each rank must perform its own full reduction, partially reduced results cannot be reused across ranks. In contrast, the ring‑based algorithm~\cite{woolley2015nccl} on the InfiniBand allows intermediate results to be forwarded and reused by downstream ranks. Due to this limitation, \nametwo achieves an average of only 1.05$\times$ relative performance compared with the InfiniBand when the message size goes beyond 256 MB. 

\textbf{Broadcast} follows a 1‑to‑N communication pattern.  %where only the root rank transmits data and all other ranks act solely as receivers. 
Relative to \name-Naive, \nameone has similar performance since \textcolor{fix}{natively \nameone uses a large data chunk size. However, \nametwo can reduce  end‑to‑end latency by 1.9$\times$ –3.6$\times$.} Such improvement stems from partitioning and distributing the root’s data across all CXL devices, allowing the root to write without interfering with concurrent read operations issued by other ranks. During the read phase, the ranks access disjoint CXL devices by varying their initial data‑chunk offsets in the memory pool. Compared with the InfiniBand interconnect, \nametwo reduces latency by  1.3$\times$ –2.8$\times$.

\textbf{Reduce} follows an N-to‑1 pattern, with the root rank performing the reduction while all other ranks write their data into the memory pool. Relative to \name-Naive, \nameone and \nametwo reduce end‑to‑end latency by 0.7$\times$–1.9$\times$ and 2.2$\times$–3.7$\times$, respectively. Compared with the InfiniBand, \nametwo achieves an additional 1.3$\times$–2.6$\times$ speedup.

\textbf{All-Gather} follows an N‑to‑N communication pattern. Compared with \name-Naive, \nameone and \nametwo reduce end‑to‑end latency by 0.7$\times$–2.9$\times$ and 1.8$\times$–5.1$\times$, respectively. \nametwo consistently achieves the highest performance because it eliminates most concurrent read/write operations targeting the same CXL device, and its fine‑grained interleaving strategy reduces the waiting time experienced by each rank.

Relative to the InfiniBand interconnect, \nametwo achieves 1.01$\times$–1.6$\times$ speedup. This improvement stems from the CXL memory pool incorporating a sufficiently large number of CXL devices, providing ample aggregate bandwidth to all participating ranks.

\textbf{Gather} follows an N-to‑1 communication pattern. %where only the root rank requires data from all other ranks. 
With the interleaving policy and fine‑grained overlapping, \nametwo achieves 4.2$\times$–8.0$\times$ speedup over \name-Naive and 1.1$\times$–2.7$\times$ speedup over the InfiniBand.

\textbf{ReduceScatter.} Compared with AllReduce, ReduceScatter requires only a subset of data from each rank. Figure \ref{fig:reduce_scatter_cxl} illustrates the workflow: unlike AllReduce, each rank in ReduceScatter processes only its assigned portion rather than the full dataset from all other ranks. With this reduced data requirement, \nametwo achieves 1.0$\times$–4.4$\times$ speedup  over \name-Naive and 0.48$\times$–1.9$\times$  speedups relative to the InfiniBand. \textcolor{camera}{When the message size is small, we found that \nametwo perform worse than InfiniBand due to  
software overheads such as cudaMemcpy invocation and synchronization. \nametwo will use fine-grained chunks to transfer the message. For example, with message size equals to 1 MB, each individual transfer is much smaller than 1 MB. In this case, each rank only needs $1/n Rank$ MB data from every other rank. Then for each data block ($1/nRank$ MB), \nametwo will split it into fine-grained data chunks. In this regime, software overheads such as cudaMemcpy invocation and synchronization become more pronounced. This hurts performance, compared with InfiniBand that relies on a more GPU-driven communication path. However, as the message size increases, the transfer granularity grows, so the software overhead is quickly amortized. Consequently, our approach achieves better performance than InfiniBand for larger message sizes. This trend is similarly visible in Scatter and AllToAll.
}

\textbf{Scatter} implements a 1‑to‑N pattern. %in which the root rank writes data into the memory pool and each remaining rank reads its  portion into a local buffer. 
Leveraging the interleaving and fine‑grained overlapping, \nametwo delivers 1.16$\times$–4.2$\times$ speedup \textcolor{dong}{over \nameone} and 0.46$\times$–1.53$\times$ speedup over the InfiniBand.

\textbf{AllToAll} follows an N-to‑N communication pattern. Its workflow is similar to ReduceScatter, with the key difference that AllToAll does not perform any reduction. Each rank exchanges distinct data segments with every other rank. With this pattern, \nametwo achieves 1.0$\times$–4.3$\times$ speedup over \name-Naive and 0.7$\times$–1.9$\times$ speedup relative to the InfiniBand.
\vspace{-5pt}
\subsection{Scalability Evaluation} 
\begin{figure*}[!ht]
    \centering
    % Row 1
    \begin{subfigure}{0.31\textwidth}
        \centering
        \includegraphics[width=\linewidth]{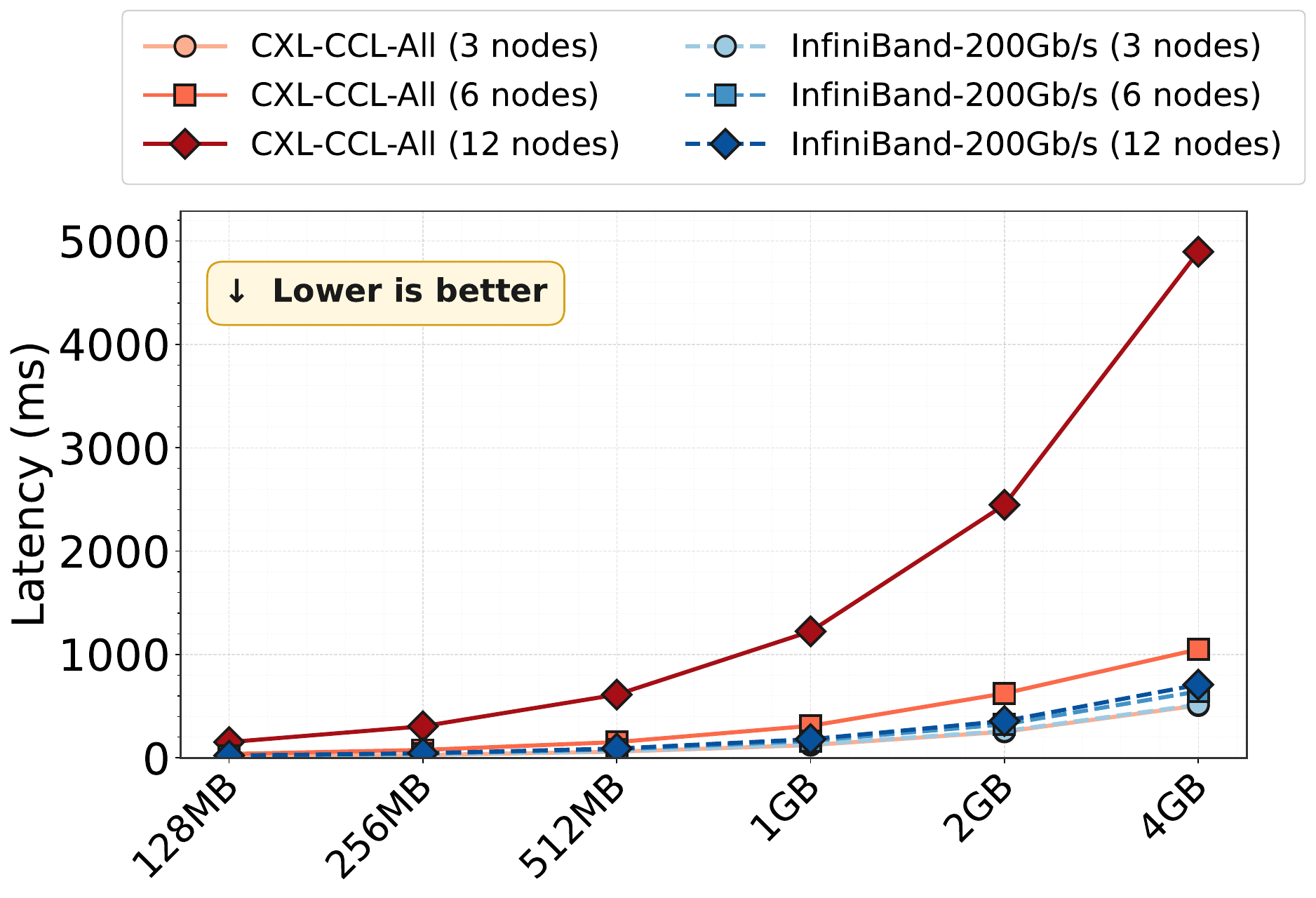}
        \caption{AllReduce.}
    \end{subfigure}
    \hspace{0.02\textwidth}
    \begin{subfigure}{0.31\textwidth}
        \centering
        \includegraphics[width=\linewidth]{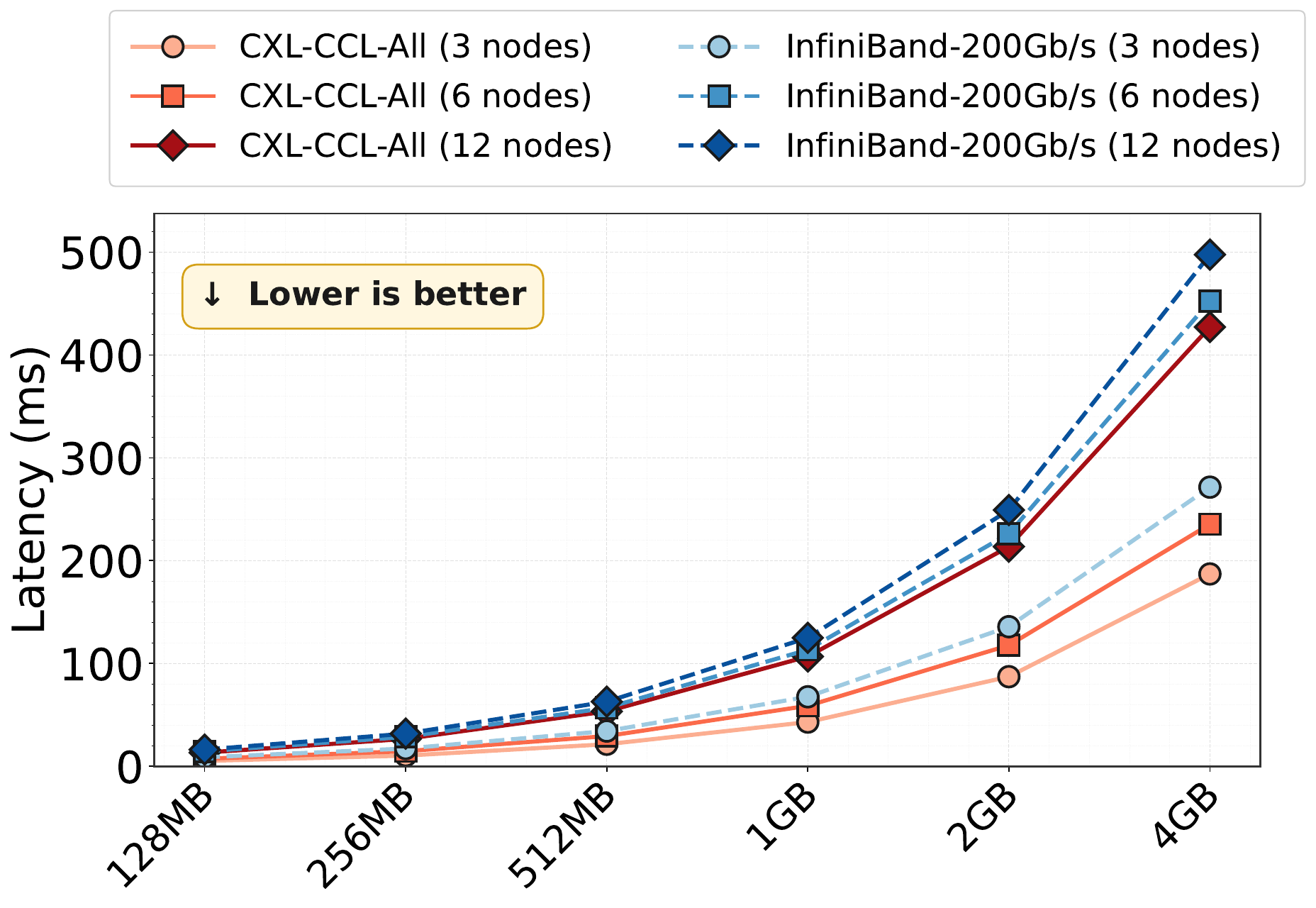}
        \caption{Broadcast.}
    \end{subfigure}
    \hspace{0.02\textwidth}
    \begin{subfigure}{0.31\textwidth}
        \centering
        \includegraphics[width=\linewidth]{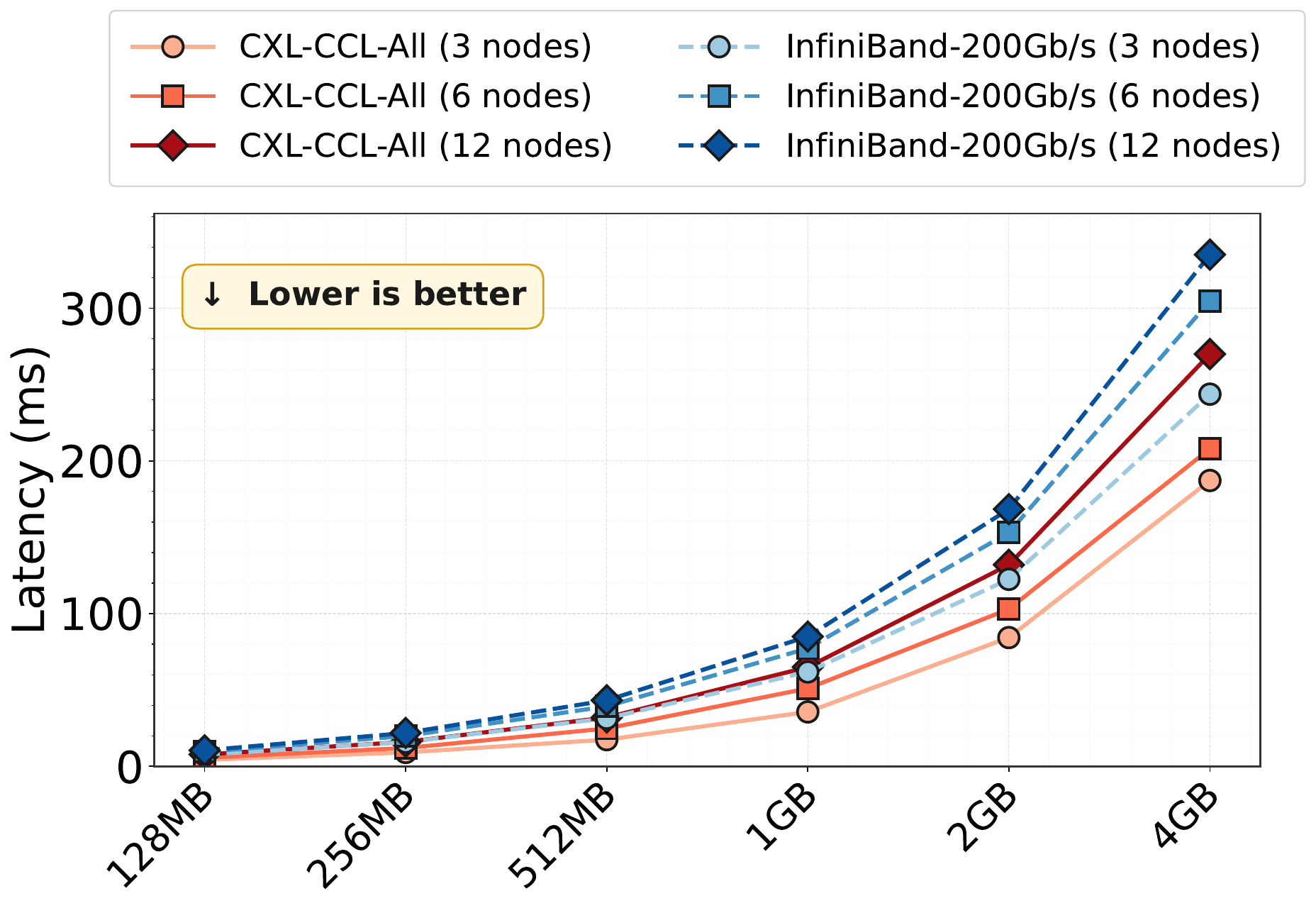}
        \caption{AllToAll.}
    \end{subfigure}

    \vspace{1em}
    
    \vspace{-20pt}
    \caption{Scalability evaluation using the CXL shared memory pool.}
    \label{fig:scale}
    \vspace{-10pt}
\end{figure*}

We evaluate four representative primitives for study. We scale the number of nodes from 3, 6 to 12, while the message size varies from 128 MB to 4 GB.  %since smaller message sizes impede accurate emulation. 
Throughout all experiments, we employ 6 CXL devices in the memory pool. \textcolor{dong}{We do not scale the system to a larger number of nodes (e.g., tens or even hundreds of nodes), because the CXL shared memory pool is expected to be shared within a small number of nodes in a data center~\cite{li2023pond, yang2025beluga,cxldb}.} Figure~\ref{fig:scale} shows the results.  

In AllReduce, each rank must access the data contributed by all other ranks to perform its local reduction. As the number of ranks increases, the volume of data that must be read from the shared memory pool grows proportionally. When the system scales from three to six servers, the execution time increases by 2.1$\times$–3.0$\times$ because each rank must read roughly 2.5$\times$ more data. Additional contention also emerges due to concurrent read and write operations targeting the same set of devices, as the CXL shared memory pool contains only six CXL devices.

When the system is further expanded to twelve nodes, the total execution time rises by 8.7$\times$–12.2$\times$. This substantial increase is primarily due to the fact that each rank must read data from other eleven ranks, significantly amplifying communication and memory‑access overhead. Moreover, with all twelve nodes simultaneously issuing both read and write requests, creating contention on shared resources becomes unavoidable and increasingly severe. \textcolor{camera}{ Compared with InfiniBand, the current NCCL implementation can eliminate redundant reductions and data transfers within the ring, which allows NCCL over InfiniBand in this scenario to achieve better scalability. }

In Broadcast, only the root rank performs write operations. When scaling to six nodes, adjusting the initial read‑chunk identifiers allows the read workload to be distributed across different devices for each rank. However, until the root completes its writes, one device inevitably experiences concurrent read and write traffic, since all writes originate from the root. Under this configuration, total execution time increases to 1.26$\times$–1.40$\times$ of the three‑node setup. When scaling to twelve nodes, the execution time grows to approximately 2.5$\times$ of the three‑node configuration. \textcolor{camera}{Compared to InfiniBand, \nametwo achieves an average speedup of approximately 1.54× across all cases.}

In AllToAll, each rank performs symmetric I/O, reading and writing an equivalent amount of data. Unlike AllReduce, AllToAll requires only a subset of each rank’s data rather than a global aggregation. \textcolor{camera}{For instance, if the message size is N bytes, then each rank will receive $N/\text{nRanks}$ bytes from every rank (including itself). Consequently, when you increase the number of nodes to 6 or 12, the total data traffic remains unchanged for the same message size N. Nevertheless, increasing the number of nodes leads to more simultaneous read/write operations on our limited CXL devices.}
As the system scales to six and twelve nodes, the end‑to‑end latency increases by 1.11$\times$–1.43$\times$ and 1.44$\times$–1.83$\times$, respectively. 

\vspace{-5pt}
\subsection{Sensitivity Study}
%We conduct a sensitivity study on the number of data chunks we should use when we do the data partitioning. We use All-Gather to show the results. We choose 256MB, 512MB and 1GB as the message size.

We conduct a sensitivity study to evaluate how the number of data chunks used during partitioning affects performance. Using the All‑Gather primitive as a representative case and set the message size equals  1 GB to illustrate the impact of different chunk configurations.

\begin{figure}[!t]
    \centering
     \includegraphics[width=0.85\linewidth]{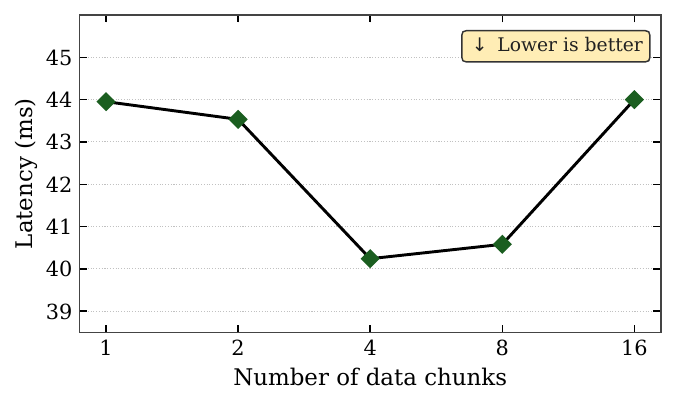}
    \caption{Sensitivity study for end-to-end latency with respect to the number of data chunks.}
    \label{fig:sensi}
    \vspace{-15pt}
\end{figure}

\textcolor{camera}{Figure \ref{fig:sensi} presents the results obtained by varying the slicing factor, i.e., the number of chunks into which the data is partitioned. For each message size, we observe that the performance changes as a function of the slicing factor. The maximum performance variation reaches approximately 9\%. The configurations with very few chunks (e.g., a single chunk) yield the worst performance. This degradation arises because large chunks strengthen data dependencies between sender and receiver ranks, reducing the degree of overlap between communication and computation. As a result, the system is unable to fully utilize the available bidirectional bandwidth of the CXL memory devices, which in turn degrades overall performance. In our tests, using 4 or 8 as the number of chunks is good for performance.}

\subsection{Using \name for LLM Training}

We evaluate \name using a real-world LLM training scenario. Specifically, \textcolor{dong}{given the GPU memory limitation,} we train a Llama-3-8B model on the dataset Wikipedia~\cite{wikidump}, using PyTorch’s Fully sharded Data Parallelism (FSDP) to perform multi-GPU training. %The results are summarized in \textcolor{red}{Table~\ref{tab:ib-cxlnccl}}, where we report normalized speedup with respect to the baseline RDMA implementation over InfiniBand.

\begin{comment}
\begin{table}[t!]
\centering
\caption{Comparison of RDMA oover InfiniBand and CXL-NCCL.}
\label{tab:ib-cxlnccl}
\begin{tabular}{|l|c|c|}
\hline
\textbf{} & \textbf{RDMA-InfiniBand} & \textbf{CXL-NCCL} \\
\hline
Normalized Speedup(s) & 1 & 1.11 \\
\hline
\end{tabular}
\end{table}
\end{comment}

FSDP uses \texttt{AllGather} and \texttt{ReduceScatter} to shuffle model parameters and to reduce gradients across distributed nodes. Using \nametwo, we achieve 1.11$\times$ speedup, compared with the baseline (RDMA over InfiniBand). Furthermore, considering the cost of constructing the interconnect, using the CXL shared memory pool costs 2.75$\times$ less than the InfiniBand, since an InfiniBand switch (supporting 200 Gbps for each port) costs \$16K, and the CXL switch costs only \$5.8K~\cite{yang2025beluga}.

%\textbf{fix}{Each subfigure is a placeholder. need to have larger font size.}

\section{Related Work}

Prior work has extensively investigated CXL-based memory architectures and their system-level implications. 
Early studies characterized the latency, bandwidth, and coherence behavior of CXL using FPGA prototypes or early CXL hardware, providing the first quantitative understanding of this emerging interconnect \cite{li2023pond, gouk2022direct, liu2025systematic, wang2024rcmp}. Research on tiered memory systems~\cite{Memtis,TPP,tiering0.8,numafault, Agarwal2017ThermostatAP,Yan2019NimblePM,acaseforgran,adaptivepagemigration,Maruf2022MULTICLOCKDT,Raybuck2021HeMemST,li2023pond,MTM, merchan,betty,plasma,sparta,ren2021zero,MD_HM,neurips20:hm-ann,SC_18,arch_tm, unimem:sc17,ren2021sentinel,cluster17:yang,cluster17:huang,hpdc16:wu,flexmem} has also begun to examine how to achieve better performance when CXL-based memory expansions are employed.
Subsequent work demonstrated the practicality of CXL in a wide range of application domains, including in-memory databases, graph processing, and deep learning workloads \cite{guo2024cxl, ahn2022enabling, yang2025unlocking, sano2023gpu, liu2024enabling, arif2022exploiting, 10793192, ma2026, song2025attncacheacceleratingselfattentioninference, wang2024exploring, wang2026hybridadaptivetuningtiered, wang2026tierbpfpagemigrationadmission, ren2021sentinel, ren2021zero,ren2025machinelearningguidedmemoryoptimization, buffalo,betty,lobster}.

Several studies further compared CXL with RDMA-based solutions and showed that CXL can provide superior performance or complementary benefits for memory-centric workloads by offering native load/store semantics rather than network-style message operations \cite{wang2024exploring, wei2023transactional}. 
Nevertheless, most of these efforts primarily focus on CPU-oriented memory expansion, performance characterization, or capacity-oriented shared memory usage. 
They do not investigate how CXL can be leveraged to support GPU-centric collective communication, nor do they address the synchronization and data placement challenges that arise when multiple GPUs across different nodes concurrently access a shared CXL memory pool. 

More recently, a growing body of work explores exposing CXL devices as shared memory across multiple nodes. 
CXL-SHM proposes a general shared-memory substrate and demonstrates its use for distributed data structures and RDMA offload, assuming device-side atomic operations \cite{zhang2023partial}. 
Tigon studies CXL-based shared memory for in-memory databases and shows that hardware coherence support is limited in scale and cannot be extended to the full device capacity \cite{huang2025tigon}. 
cMPI replaces the traditional point-to-point network path in MPI communication with CXL shared memory to enable inter-node message passing \cite{wang2025cmpi}. 
These systems primarily treat CXL as a general shared-memory substrate or as an alternative communication medium for distributed applications. In contrast, this work focuses on enabling efficient GPU collective communication on top of a shared CXL memory pool. \name directly exploits the memory semantics provided by CXL to coordinate data exchange among GPUs and explicitly addresses challenges that are unique to collective operations, including the lack of cross-node synchronization, the absence of hardware-managed fine-grained interleaving across multiple CXL devices, and the need to overlap data publication and consumption to reduce end-to-end latency.

\vspace{-10pt}
\section{Conclusions}
Efficient collective GPU communication is essential for high-performance AI workloads. Different from the traditional approaches that rely on RDMA and high-performance interconnect, we introduce a fundamentally new method that leverages a CXL shared memory pool for GPU collective communication. Based upon the emerging CXL-based architecture and memory copy-based communication implementation, we demonstrate the superior performance of our approach. When compared with a 200 Gbps InfiniBand, our evaluation results indicate that \name achieves large speedup with various message sizes.
%1.34$\times$ speedup for AllGather, 1.84$\times$ for Broadcast, 1.94$\times$ for Gather, 1.04$\times$ for Scatter, 1.5$\times$ for AllReduce, 1.43$\times$ for ReduceScatter, 1.70$\times$ for Reduce, and 1.53$\times$ for AlltoAll, averaged over varying message sizes.


\bibliographystyle{ACM-Reference-Format}
\bibliography{li, han,dong}


% \appendix








\end{document}